\documentclass[twocolumn,showpacs,preprintnumbers,amsmath,amssymb,APSl,prd,nofootinbib,superscriptaddress]{revtex4-1}

\usepackage{dcolumn}% Align table columns on decimal point
\usepackage{bm}% bold math
\usepackage{ifpdf}
\usepackage{hyperref}
\usepackage{dcolumn}
\usepackage{bm}
\usepackage[spanish,english]{babel}
\usepackage{amsfonts}
\usepackage{amssymb}
\usepackage{graphicx}
\usepackage{color}
\usepackage[latin1]{inputenc}

\newcommand \be{\begin{equation}}
\newcommand \en{\end{equation}}
\newcommand \bea{\begin{eqnarray}}
\newcommand \ena{\end{eqnarray}}

\begin{document}

\title{What is a singular black hole beyond General Relativity?}

\author{Cecilia Bejarano} \email{cbejarano@iafe.uba.ar}
\affiliation{Instituto de Astronom\'ia y F\'isica del Espacio (IAFE, CONICET-UBA),
Casilla de Correo 67, Sucursal 28, 1428 Buenos Aires, Argentina.}
\affiliation{Departamento de F\'{i}sica Te\'{o}rica and IFIC, Centro Mixto Universidad de
Valencia - CSIC. Universidad de Valencia, Burjassot-46100, Valencia, Spain.}
\author{Gonzalo J. Olmo} \email{gonzalo.olmo@uv.es}
\affiliation{Departamento de F\'{i}sica Te\'{o}rica and IFIC, Centro Mixto Universidad de
Valencia - CSIC. Universidad de Valencia, Burjassot-46100, Valencia, Spain.}
\affiliation{Departamento de F\'isica, Universidade Federal da
Para\'\i ba, 58051-900 Jo\~ao Pessoa, Para\'\i ba, Brazil.}
\author{Diego Rubiera-Garcia} \email{drgarcia@fc.ul.pt}
\affiliation{Instituto de Astrof\'{\i}sica e Ci\^{e}ncias do Espa\c{c}o, Faculdade de
Ci\^encias da Universidade de Lisboa, Edif\'{\i}cio C8, Campo Grande,
P-1749-016 Lisbon, Portugal.}


\date{\today}

\begin{abstract}
Exploring the characterization of singular black hole spacetimes, we study the relation between energy density, curvature invariants, and geodesic completeness using a quadratic $f(R)$ gravity theory coupled to an anisotropic fluid. Working in a metric-affine approach, our models and solutions represent minimal extensions of General Relativity (GR) in the sense that they rapidly recover the usual Reissner-Nordstr\"{o}m solution from near the inner horizon outwards. The anisotropic fluid helps modify only the innermost geometry. Depending on the values and signs of two parameters on the gravitational and matter sectors, a breakdown of the correlations between the finiteness/divergence of the energy density, the behavior of curvature invariants, and the (in)completeness of geodesics is obtained. We find a variety of configurations with and without wormholes,  a case with a de Sitter interior, solutions that mimic non-linear models of electrodynamics coupled to GR, and configurations with up to four horizons. Our results raise questions regarding what infinities, if any, a quantum version of these theories should regularize.
\end{abstract}

\maketitle

\section{Introduction}

One of the most serious drawbacks associated to Einstein's theory of General Relativity (GR) is the unavoidable existence, under reasonable physical assumptions, of spacetime singularities deep inside black holes, as well as in the early universe \cite{Theorems}. This is due to the fact that at such singularities the predictability of physical laws comes to an end because measurements are no longer possible. The underlying reason is that the existence of incomplete geodesics implies the destruction/creation of observers and/or information (light signals) as some limiting boundaries are approached.  As a way out of this problem, Penrose introduced \cite{CCC} the cosmic censorship conjecture, by which singularities emerging out of gravitational collapse are assumed to be hidden behind an event horizon, so they cannot causally affect physical processes taking place in the portion of universe accessible to far away observers. Since sweeping the problem under the carpet does not solve it,  finding a consistent description of the interaction between gravity and matter, where the resolution of spacetime singularities may be naturally achieved, has become a major goal from different perspectives (classical and quantum, fundamental and phenomenological).

It is typically argued that spacetime singularities should be resolved by a quantum theory of gravity. This is supported by the idea that the quantum degrees of freedom of the gravitational field are expected to be non-negligible in regions of very high curvature. This view, inherited from the effective field theory approach to quantum theory, is very appealing but should be taken with care in gravitational scenarios, where the notion of singularity is not necessarily tied to the divergence of some quantities in some regions \cite{Geroch,Wald,Hawking-Ellis}. For geometric theories of gravity (classical theories), the very existence of observers is more fundamental than the possibility of obtaining absurd results in a measurement, as the latter is not possible without the former. It is for this reason that the existence of incomplete geodesics, for which the affine parameter is not defined over the whole real line, appears as the key element in the singularity theorems.

In the context of GR, the incompleteness of geodesics usually occurs simultaneously with the divergence of scalar quantities, such as the energy density of the matter sources or certain curvature invariants. These divergences appear as a {\it reason} for the incompleteness of the geodesics, leading to a {\it rule of thumb} for the identification of singular spacetimes \cite{phy} (see \cite{HS} for a critical viewpoint on this issue). Indeed this has shaped many approaches to the singularity problem based on the idea that such quantities should remain bounded (see e.g. \cite{Ansoldi} for a review).

One of such approaches is given by classical non-linear models of the electromagnetic field. This is supported on the success of Born-Infeld theory of electrodynamics, where a square-root modification of the Maxwell action gets rid of the divergence of the self-energy of Coulomb's field by imposing a maximum bound on the electric field at the center \cite{BI}. It is natural to wonder whether a similar mechanism for the removal of singularities could occur in the context of gravitation. In this sense Born-Infeld electrodynamics, though successful in making the energy density of the electromagnetic field finite, fails to keep at bay divergences on the curvature scalars when coupled to gravity, which comes alongside with the incompleteness of (some) geodesics \cite{BI-grav}. In this regard, similar attempts using other well defined non-linear electrodynamics models have failed as well \cite{NED-grav}. Nonetheless, it is worth mentioning that some examples of non-linear electrodynamics do regularize curvature divergences \cite{AB}, but such models are constructed in an ad hoc way and yield unphysical features, as shown by Bronnikov \cite{Bronnikov} (see also \cite{Novello}). This strategy has been extended to the case of gravitational actions going beyond the Einstein-Hilbert Lagrangian of GR, such as Gauss-Bonnet and, more generally, Lovelock theories \cite{GB}, where similar disappointing results have been obtained (see e.g. \cite{GBNED} for some attempts in this context). Consequently, it is fair to say that such models have been unable to find a fully consistent way out of the singularity problem in GR.

In this work we shall examine in detail the relation between energy density, curvature invariants, and geodesic completeness in some theories of gravity beyond GR. This will allow us to see if the correlations observed in GR among those quantities still persist in other gravitational theories (see \cite{RegularBGR} for related ideas explored in this context). In other words, can matter/curvature infinities be seen as the {\it reason} for the incompleteness of geodesics? This study is relevant in order to understand what problems, if any, a quantum version\footnote{Note that we are assuming that any classical theory of gravity should admit a quantum version. } of those theories of gravity should solve.

In our approach, we interpret gravitation as a geometric phenomenon, but geometry as something more than just curvature. In the metric-affine (or Palatini) formulation of classical gravitation, geometric properties such as non-metricity and torsion, besides curvature, are allowed by construction. The lack of these freedoms in the usual Riemannian approach could be an excessive constraint with a potentially non-negligible impact on the problems that gravity theories typically exhibit at high-energy. It should be noted that non-metricity and torsion are necessary to deal with different kinds of geometric defects in continuum systems with a microstructure, such as Bravais crystals or graphene \cite{SSP}. For this reason, metric-affine geometry is commonly used in the study of condensed matter physics \cite{PL}. Nonetheless, for operational convenience, in this work we shall neglect torsion (see, however, \cite{ORtorsion} for a discussion on the role of torsion in metric-affine theories) and focus on non-metricity only \cite{Pal}. Indeed, the question of whether gravity as a manifestation of the curvature\footnote{As a matter of fact, gravity could be interpreted as a manifestation of torsion in a flat background, such as in the teleparallel formulation of general relativity (see e.g. \cite{TEGR}), but also it could belong to a more general picture where curvature and torsion are both required to properly describe the gravitational interaction as in the case of Einstein-Cartan theories \cite{EC}.} of spacetime is purely a matter of metrics or if the affine structure of spacetime is on equal footing as the metric one has been at debate since soon after the establishment of GR (see e.g. \cite{Zanelli} for a pedagogical discussion). Certainly, when GR is formulated \`a la Palatini, the variation of the action with respect to the independent connection yields a set of equations that simply express the metric-connection compatibility condition. The fact that this approach yields the same dynamics as that of considering the metric as the only independent degree of freedom (\emph{metric} approach) has frequently lead to regard the Palatini variation as merely an alternative way to deriving the field equations of GR. For other theories of gravity, however, the compatibility between metric and connection is broken and the peculiarities of the metric-affine approach become manifest.

The scenario considered here corresponds to a simple quadratic $f(R)$ gravity extension of GR (for which many applications have been investigated in the literature, see e.g. \cite{fRlit}), formulated in a metric-affine framework. It should be pointed out that with the advent of the gravitational wave astronomy following the discovery of GW150914 by LIGO \cite{LIGO}, both gravitational extensions of GR and exotic compact objects in such models can be put to experimental test \cite{ECOs}. As the matter sector, in our setup we consider an anisotropic fluid (constrained to satisfy standard energy conditions), which has been recently investigated in some detail in a number of astrophysical/cosmological scenarios \cite{af}. Such fluids include a number of particularly interesting cases, such as that of non-linear electrodynamics. The resulting spacetimes are split into four different cases, depending on the combinations of the signs of the coupling constant of the quadratic gravity contribution and of a constant associated to the matter sector. A noteworthy feature of many of the solutions obtained is the emergence of a finite-size wormhole structure [see \cite{Visser} for detailed account on wormhole physics] replacing the point-like singularity typically found at the center of GR black holes. It is worth pointing out that wormholes have been suggested as solutions to spacetime singularities in approaches to quantum gravity such as loop quantum gravity \cite{LQG} and shape dynamics \cite{SD} (see also \cite{BroFab} and references therein, where wormholes are linked to regularization mechanisms.)

The main aim of the present work is to determine when the typically assumed correlation between divergence of curvature scalars and geodesic incompleteness is broken. In this sense, we note that the concept underlying the formulation of the singularity theorems \cite{Wald} is that of geodesic completeness, namely, whether a geodesic curve can be extended to arbitrarily large values of its affine parameter or not. This is a logically independent and more primitive concept than that of curvature divergences [see \cite{phy} for a nice discussion on this issue], with the latter playing no role on such theorems. As already mentioned, the widespread identification between them in the literature is explained as due to the fact that in many cases of interest (particularly in GR) those spacetimes having (some) incomplete geodesics, also yield (some) divergent curvature scalars \cite{Ansoldi}. In some of the spacetimes found here we explicitly show that the presence of wormholes yield geodesically complete spacetimes, though curvature scalars may blow up at the wormhole throat. In other cases without wormholes, we meet the incompleteness of geodesics despite the finiteness of curvature scalars. The relation of these magnitudes with the (boundedness of the) energy density of the matter fields is also discussed.

The paper is organized as follows: in Sec. \ref{sec:II} we introduce the action and main equations of $f(R)$ gravity formulated \`{a} la Palatini. In Sec. \ref{sec:III} we specify the matter sector of our theory under the form of an anisotropic fluid and introduce a number of constraints on it. Next, in Sec. \ref{sec:IV}, we focus our discussion upon a quadratic $f(R)$ model and solve the field equations for the metric. Sec. \ref{sec:V} contains the main results of this work, where we study the four different classes of spacetimes, and discuss in detail the relation between energy density, curvature scalars, and geodesic completeness. We conclude in Sec. \ref{sec:VI} with a summary and some perspectives.

\section{Action and main equations} \label{sec:II}

The action of $f(R)$ gravity can be written as

\begin{equation}  \label{eq:action}
S=\frac{1}{2\kappa^2} \int d^4 x \sqrt{-g} f(R) + S_m(g_{\mu\nu},\psi_m) \ ,
\end{equation}
with the following definitions and conventions: $\kappa^2$ is Newton's constant in suitable units (in GR, $\kappa^2= 8 \pi G/c^4$), $g$ is the determinant of the spacetime metric $g_{\mu\nu}$,  $f(R)$ is a given function of the curvature scalar, $R \equiv g_{\mu\nu} R^{\mu\nu}$, where the Ricci tensor, $R_{\mu\nu} \equiv R_{\mu\nu}(\Gamma)$, which follows from the Riemann tensor as $R_{\mu\nu} \equiv {R^\alpha}_{\mu\alpha\nu}$, is entirely built out of the affine connection, $\Gamma \equiv \Gamma^{\lambda}_{\mu\nu}$, which is a priori independent of the metric (metric-affine or Palatini approach). Finally, $S_m$ is the matter action, which is assumed to depend only on the matter fields, collectively denoted as $\psi_m$, and on the metric $g_{\mu\nu}$.

Performing independent variations of the action (\ref{eq:action}) with respect to metric and connection one gets two systems of equations
\begin{eqnarray}
f_R R_{\mu\nu}-\frac{f}{2} g_{\mu\nu}&=&\kappa^2 T_{\mu\nu} \label{eq:metric} \ ,\\
\nabla_{\lambda}^{\Gamma}(\sqrt{-g} f_R g^{\mu\nu})&=&0 \label{eq:connection} \ ,
\end{eqnarray}
where $f_R \equiv df/dR$ and $T_{\mu\nu}=-\frac{2}{\sqrt{-g}} \frac{\delta S_m}{\delta g^{\mu\nu}}$ is the stress-energy tensor of the matter. It is worth mentioning that Eq.~(\ref{eq:connection}) simply states that the independent connection fails to be metric or, in other words, that a non-metricity tensor ${Q_{\lambda\mu\nu}} \equiv \nabla_{\lambda}^{\Gamma}g_{\mu\nu} \neq 0$ is present. In the GR case, $f_R=1$ and Eq.~(\ref{eq:connection}) becomes $\nabla_{\lambda}^{\Gamma}(\sqrt{-g} g^{\mu\nu})=0$, which is fully equivalent to $\nabla_{\lambda}^{\Gamma}g_{\mu\nu} = 0$  and thus $\Gamma_{\mu\nu}^{\lambda}$ becomes the Levi-Civita connection of the metric $g_{\mu\nu}$, while the field equations (\ref{eq:metric}) boil down to those of GR with possibly a cosmological constant term. This is the underlying reason for the equivalence between the Palatini and metric formulations of GR. For more general $f(R)$ Lagrangians, however, non-metricity becomes an inherent feature of the field equations.

It is also important to understand the intimate relation existing between matter and gravity in Palatini theories of gravity. Tracing with $g^{\mu\nu}$ in Eq.~(\ref{eq:metric}) yields the result
\begin{equation} \label{eq:trace}
Rf_R-2f=\kappa^2 T \ ,
\end{equation}
where $T$ is the trace of the stress-energy tensor. This is not a differential equation, but instead it just establishes an algebraic, non-linear relation between curvature and matter. Given an $f(R)$ theory, solving Eq.~(\ref{eq:trace}) yields a solution $R=R(T)$, which generalizes the GR relation, $R=-\kappa^2 T$. This algebraic relation explains the absence of extra dynamical degrees of freedom in our theory as compared to the usual metric approach, where the scalar curvature satisfies a second-order differential equation, thus implying the presence of propagating scalar degrees of freedom. In the Palatini case, the additional curvature terms are just nonlinear functions of $T$ and can be collected as extra pieces in an effective stress-energy tensor. This way, the Palatini field equations for the metric (\ref{eq:metric}) can be simply written as
\begin{equation} \label{eq:GRlike}
G_{\mu\nu}=\kappa^2 \tau_{\mu\nu} \ ,
\end{equation}
where the \emph{effective} stress-energy tensor is written as
\begin{eqnarray}
\tau_{\mu\nu}&=&\frac{\kappa^2}{f_R}T_{\mu\nu}-\frac{Rf_R-f}{2f_R}g_{\mu\nu}\nonumber \\ &-& \frac{3}{2f_R^2}\left[\partial_\mu f_R\partial_\nu f_R-\frac{1}{2}g_{\mu\nu}(\partial f_R)^2\right]\nonumber\\ &+&\frac{1}{f_R}\left[\nabla_\mu \nabla_\nu f_R-g_{\mu\nu}\Box f_R\right] \ .
\end{eqnarray}
However, from a practical point of view, in many cases of interest it is easier to solve the field equations by noting that the result $R=R(T)$ allows us to introduce in Eq.~(\ref{eq:connection}) a rank-two tensor $h_{\mu\nu}$ satisfying
\begin{equation} \label{eq:hdef}
\nabla^{\Gamma}_{\lambda}(\sqrt{-h} h^{\mu\nu})=0 \ ,
\end{equation}
such that the independent connection $\Gamma_{\mu\nu}^{\lambda}$ can be expressed as the Christoffel symbols of the metric $h_{\mu\nu}$, i.e.,
\begin{equation}\label{eq:LCfR}
\Gamma^\lambda_{\mu\nu}=\frac{h^{\lambda\rho}}{2}\left[\partial_\mu h_{\rho\nu}+\partial_\nu h_{\rho\mu}-\partial_\rho h_{\mu\nu}\right] \ .
\end{equation}
Comparing this with Eq.~(\ref{eq:connection}), it is immediately seen that the physical metric $g_{\mu\nu}$ can be obtained out of $h_{\mu\nu}$ according to the conformal transformations
\begin{equation} \label{eq:conformal}
h_{\mu\nu} = f_R g_{\mu\nu} \quad ; \quad  h^{\mu\nu} = f_R^{-1} g^{\mu\nu} \ ,
\end{equation}
where, recall,  $f_R$ is a function of the matter, $f_R\equiv f_R(T)$.

An alternative representation of the field equations is now possible in terms of $h_{\mu\nu}$ by contracting Eq.~(\ref{eq:metric}) with $h^{\alpha\mu}$ and using the relations (\ref{eq:conformal}) to obtain
\begin{equation} \label{eq:fieldh}
{R^\mu}_\nu(h)=\frac{1}{f_R^2} \left(\frac{f}{2} {\delta^\mu}_\nu + \kappa^2 {T^\mu}_\nu \right) \ ,
\end{equation}
where ${R_{\mu\nu}}(h) \equiv {R_{\mu\nu}}(\Gamma)$ is the Ricci tensor constructed with the Christoffel symbols of the metric $h_{\mu\nu}$, see Eq.~(\ref{eq:LCfR}). Note that due to the fact that $f \equiv f(R(T))$ all the objects on the right-hand-side of Eq.~(\ref{eq:fieldh}) are just functions of the matter. Thus Eq.~(\ref{eq:fieldh}) represents a set of second-order field equations for $h_{\mu\nu}$ and, since the conformal transformations (\ref{eq:conformal}) depend only on the matter sources, the field equations for $g_{\mu\nu}$ will be second-order as well. In vacuum, ${T^\mu}_{\nu}=0$, one has $g_{\mu\nu}=h_{\mu\nu}$ (up to a trivial re-scaling of units) and the field equations (\ref{eq:fieldh}) reduce to those of GR with a cosmological constant term, which confirms the absence of ghost-like propagating degrees of freedom in these theories.

\section{Anisotropic fluids} \label{sec:III}

In this work we are interested on obtaining black hole solutions in Palatini $f(R)$ theories, and to compare their structure with that of electrically charged black holes of GR. However, due to the fact that the non-linear corrections appearing on the right-hand-side of the new gravitational field equations (either in Eq.~(\ref{eq:GRlike}) or Eq.~(\ref{eq:fieldh})) depend just on the trace of the matter, $f(R) \equiv f(R(T))$, the new dynamics encoded in Palatini $f(R)$ theories can only be excited when non-traceless stress-energy tensors are considered. This implies that considering a classical Maxwell electromagnetic field, whose trace is zero, would yield electrovacuum solutions identical to those of GR with a cosmological constant (Reissner-Nordstr\"om-Anti-de Sitter black holes). Thus, in order to explore new physics in these scenarios, we must consider stress-energy tensors with a non-vanishing trace. One can then assume that a trace anomaly or other types of corrections are generated by quantum effects and propose a stress-energy tensor of the following form:
\begin{equation} \label{eq:tani}
{T_\mu}^\nu = \text{diag}(-\rho,P_r,P_{\theta},P_{\varphi}) \ .
\end{equation}
This corresponds to an anisotropic fluid, where $\rho$ is the energy density and $\{P_r, P_{\theta}, P_{\varphi}\}$ are the (different, in principle) pressures. This class of fluids has been recently considered in Refs.~\cite{fluids1,fluids2,Tamang} where, working in slightly different scenarios,  it was found that wormhole solutions can be constructed\footnote{Here the word ``constructed" means that the wormhole geometry is given first, and then the gravitational field equations are driven back in order to find the matter sources threading the geometry. This is a widely spread strategy in the context of wormhole physics \cite{Visser}.} in Eddington-inspired Born-Infeld theories of gravity without violation of the energy conditions. In contrast to that approach, as we shall show below, in the Palatini $f(R)$ scenario considered here, wormholes can be obtained directly as solutions of the field equations without {\it a priori} designer approach.

\subsection*{Fluid model}

To simplify the analysis and obtain analytically accessible scenarios, let us constrain the functions defining our model. First we restrict the fluid to satisfy $P_r=-\rho$ and $P_{\theta}=P_{\varphi}=K(\rho)$, where $K(\rho)$ is a free input function whose form will be specified later. Thus, the stress-energy tensor for this fluid reads
\begin{equation} \label{eq:sef}
{T_\mu}^{\nu}=\text{diag}[-\rho,-\rho, K(\rho),K(\rho)] \ .
\end{equation}
A motivation for considering these constraints is the fact that the form of the stress-energy tensor (\ref{eq:sef}) exactly matches that of some non-linear theories of electrodynamics. Indeed, in such a case, defining the matter model as a given function $\varphi(X,Y)$ of the two field invariants $X=-(1/2) F_{\mu\nu}F^{\mu\nu}$ and $Y=-(1/2) F_{\mu\nu}F^{*\mu\nu}$, that can be built out of the field strength tensor $F_{\mu\nu}=\partial_{\mu}A_{\nu}-\partial_{\nu}A_{\mu}$ and its dual $F^{\mu\nu *}= \frac{1}{2} \epsilon^{\mu\nu\alpha\beta}F_{\alpha\beta}$, the corresponding stress-energy tensor is written as
\begin{equation}
{T_\mu}^{\nu}=\frac{1}{8\pi}\text{diag}[\varphi-2(X\varphi_X+Y\varphi_Y), \varphi-2(X\varphi_X+Y\varphi_Y), \varphi,\varphi] \ ,
\end{equation}
where $\varphi_X \equiv d\varphi/dX$ and $\varphi_Y \equiv d\varphi/dY$. Identifying $-8\pi \rho = \varphi-2(X\varphi_X+Y\varphi_Y)$ and $8\pi K(\rho)= \varphi$, it is clear that specifying a function $K(\rho)$ allows to solve these equations to determine the function $\varphi(X,Y)$, at least in implicit form, associated to the anisotropic fluid under consideration.

To obtain additional information on the fluid described by the stress-energy tensor (\ref{eq:sef}), using the fact that the independent connection $\Gamma_{\mu\nu}^{\lambda}$ does not couple to the matter in the action (\ref{eq:action}), one finds that the standard conservation equation, $\nabla_\mu {T^\mu}_\nu=0$, holds in these theories. Now, considering static spherically symmetric spacetimes, we can write a line element for the spacetime metric $g_{\mu\nu}$ as
\begin{equation} \label{eq:metricg}
ds^2=-C(x)dt^2+B^{-1}(x) dx^2+r^2(x)(d\theta^2+\sin^2\theta d\varphi^2) \ ,
\end{equation}
where the functions $C(x)$, $B(x)$ and $r(x)$ are to be determined by integration of the gravitational field equations.

With this line element, the  conservation equation above just reads $\rho_x+2[\rho+K(\rho)]r_x/r=0$, where $\rho_x \equiv d\rho/dx$ and $r_x \equiv dr/dx$, which can be integrated to give a relation between $r(x)$ and $\rho(x)$ as

\begin{equation} \label{eq:rrho}
r^2(x)=r_0^2 \exp\left[{-\int^\rho \frac{d\tilde{\rho}}{\tilde{\rho}+K(\tilde{\rho})}}\right] \ ,
\end{equation}
where $r_0$ is an integration constant with dimensions of length and $\tilde{\rho}$ is the energy density without dimensions. To proceed further and integrate explicitly this equation, we need to specify a function $K(\rho)$. Let us take the choice
\begin{equation} \label{eq:Kchoice}
K(\rho)=\alpha \rho + \beta \rho^2,
\end{equation}
where, for dimensional consistency, $\alpha$ is a dimensionless constant and $\beta$ has dimensions of inverse density. This choice covers a number of interesting cases and allows us to obtain analytical solutions. Indeed, in this case, from the expression (\ref{eq:Kchoice}), the relation between $\rho(x)$ and $r(x)$ in Eq.~(\ref{eq:rrho}) is explicitly written as
\begin{equation} \label{eq:ed1}
\rho(r)=\frac{(1+\alpha) \rho_0}{\left(\frac{r}{r_0}\right)^{2(1+\alpha)}-\beta \rho_0} \ ,
\end{equation}
where $\rho_0$ is a reference energy density that arises as an integration constant and can be fixed from the asymptotic behavior of the fluid. In particular, for $\alpha=1$, the fluid density and the metric far from the center tend to those generated by a Maxwell field, namely, $\rho r^4=\frac{q^2}{8\pi}$, which allows to relate $\rho_0 r_0^4$ with the electric charge, $q$. Moreover, if $\beta=0$, the stress-energy tensor of the fluid exactly becomes that of a Maxwell field with a vanishing trace and, as already mentioned, this yields the same dynamics as that of GR. However, non-trivial combinations of $\alpha$ and $\beta$ provide modified field equations and generate new solutions.

The analysis now requires to be split into the cases $\beta<0$ and $\beta>0$, since their properties are very different. For $\beta >0$ there is a critical radius $r_{\star}=(|\beta|\rho_0)^{1/[2(1+\alpha)]} r_0$ at which the energy density blows up. Thus the location of the standard divergence in the density of the fluid (Maxwell case) shifts from $r=0$ to the finite radius $r_{\star}$. On the other hand, for the case $\beta<0$ the energy density is finite everywhere, having a maximum value
\begin{equation} \label{eq:bound}
\rho_m=\frac{(1+\alpha)}{\vert \beta \vert} \ ,
\end{equation}
at the center.  This is quite a similar result as that found in certain models of non-linear electrodynamics, such as the one of Born and Infeld \cite{BI}, where the electric field attains a maximum value at the center and regularizes the energy density. In Sec.\ref{sec:V} we will study the implications and impact of the finiteness (or not) of the energy density, via the bound (\ref{eq:bound}), on the regularity of the corresponding spacetimes. Note in this sense that the particular case with $\beta=0$ and $0<\alpha<1$ was studied in detail in Ref.\cite{Universe}.

To simplify the analysis and the notation let us fix $\alpha=1$ from now on and define $\tilde{\beta}=s_\beta|\beta| \rho_0$, with $s_\beta=\pm 1$ denoting the sign of $\beta$, and introduce the dimensionless variable $z=r/r_{\star}$, with $r_{\star}$ the critical radius defined above. Then, we get  $z^4=s_{\beta} r^4/\tilde{\beta} r_0^4$ so that the energy density of the fluid simply reads
\begin{equation} \label{eq:rhoz}
\rho=\frac{\rho_m}{z^4-s_\beta} \ .
\end{equation}

To conclude this section, we emphasize that we are only considering   matter sources satisfying the energy conditions. For instance, the weak energy condition (WEC) states that the following conditions have to be fulfilled \cite{Visser}: $\rho>0$ and $\rho + p_i>0$ ($i=r,\theta,\varphi$) in Eq.~(\ref{eq:tani}). For the particular ansatz (\ref{eq:sef}) with the choice (\ref{eq:Kchoice}) and the expressions for the energy density (\ref{eq:ed1}) and (\ref{eq:bound}), it follows that the WEC will be satisfied whenever $\alpha>0$, which is consistent with the choice $\alpha=1$ above.

\section{Gravity model and formal solutions} \label{sec:IV}

To work with the simplest possible scenario, let us consider the quadratic $f(R)$ model
\begin{equation}
f(R)=R-\sigma R^2 \ ,
\end{equation}
where $\sigma$ is a constant with dimensions of length squared. This model is particularly amenable for calculations because the trace equation (\ref{eq:trace}) yields $R=-\kappa^2 T$, which is the same linear relation as in GR, this result being just an accident related to the functional form of the quadratic model in four dimensions. With this choice, we find that the quantity $f_R$, which will play a key role in the characterization of the solutions, takes the simple form\footnote{If in the gravity Lagrangian we allow $\sigma$ to take positive and negative values, then $\gamma$ should be parameterized as $s_\gamma |\gamma|$. This leads to four types of models depending on the different combinations of $s_\beta$ and $s_\gamma$.}
\begin{equation} \label{eq:fz}
f_R=1+s_\beta s_\sigma \frac{\gamma }{(z^4-s_\beta)^2} \ ,
\end{equation}
where $\gamma \equiv \rho_m/\rho_{\sigma}$ (and we have introduced $\rho_{\sigma} \equiv 1/(8\kappa^2 |\sigma|)$ to denote the energy scale associated to the gravitational coupling constant $\sigma=s_\sigma |\sigma|$) represents the relative strength between the matter and gravitational sectors, such that the GR limit is recovered when $\gamma\to 0$. Note that the parametrization of $\sigma$ with $s_\sigma$ and of $\beta$ with $s_{\beta}$ leads to four different configurations, which will be studied separately in Sec.\ref{sec:V} .

\subsection{The metric} \label{subsec:sol}

To solve the field equations (\ref{eq:fieldh}) we introduce a static, spherically symmetric line element for the auxiliary metric $h_{\mu\nu}$ as
\begin{equation} \label{eq:metrich}
ds_h^2= -e^{2\Phi(x)}A(x)dt^2 + \frac{1}{A(x)} dx^2 + x^2 (d\theta^2+\sin^2\theta d\varphi^2)\ ,
\end{equation}
where $\Phi(x)$ and $A(x)$ are two functions to be determined using the field equations (\ref{eq:fieldh}). From the symmetry ${T^t}_t={T^x}_x$ one finds that  ${R^t}_t-{R^x}_x=0$, which implies that $\Phi(x)=$ constant, which can be put to zero by a redefinition of the time coordinate without loss of generality. The remaining field equation follows from the component
\begin{equation} \label{eq:Rthth}
{R^\theta}_\theta(h)= \frac{1}{x^2}(1-A -x A_x) \ ,
\end{equation}
which can be simplified by introducing the mass ansatz
\begin{equation} \label{eq:Ax}
A(x)=1-\frac{2M(x)}{x} \ ,
\end{equation}
leading to the first-order equation
\begin{equation} \label{eq:thth}
2\frac{M_x}{x^2} =\frac{1}{f_{R}^2}\left( \frac{f}{2}+\kappa ^2 {T^{\theta}}_\theta \right)\ ,
\end{equation}
where $M_x \equiv dM/dx$. To handle the integration of the mass function $M(x)$ it is useful to take a parametrization
\begin{equation} \label{eq:Mz}
M(x)=M_0(1+\delta_1 G(x)) \ ,
\end{equation}
with $2 M_0 \equiv r_S$ representing the Schwarzschild radius and $\delta_1$ a dimensionless constant defined as
\begin{equation}
\delta_1\equiv \frac{\kappa^2\rho_m (r_0|\tilde{\beta}|^{\frac{1}{4}})^{3}}{r_S}  \ .
\end{equation}
This puts forward that $M(x)$ is made out of a constant contribution, $M_0$, plus a term generated by the fluid and represented by the function $G(x)$ (see Eq. (\ref{eq:GzfR}) below). The resulting solution allows to construct the physical metric $g_{\mu\nu}$ by means of the conformal relations (\ref{eq:conformal}). This way, the physical line element  can be written as
\begin{equation} \label{eq:lineel}
ds^2=-\frac{A(x)}{f_R} dt^2 + \frac{dx^2}{A(x)f_R} + r^2(x) (d\theta^2+\sin^2\theta d\varphi^2) \ .
\end{equation}
Taking now into account that such conformal transformations also imply that
\begin{equation} \label{eq:xr}
x^2=f_R(r) r^2 \ ,
\end{equation}
whose dimensionless version using $z$ and $\tilde{x}\equiv x/r_\star$ is
\begin{equation} \label{eq:tildexz}
\tilde{x}^2=f_R(z) z^2 \ ,
\end{equation}
and then we obtain the relation
\begin{equation}
\frac{dz}{d\tilde{x}}= \frac{1}{f_R^{1/2} \left[1+\frac{1}{2} \frac{z f_R,z}{f_R} \right]} \ ,
\label{eq:dzdx}
\end{equation}
which allows us to express (\ref{eq:thth}), by means of Eq.~(\ref{eq:Mz}), as a differential equation involving only the variable $z$:
\begin{eqnarray}\label{eq:GzfR}
G_{z}&=&\frac{z^2}{(z^4-s_\beta)f_R^{3/2}}\left(1-\frac{s_\sigma\gamma}{(z^4-s_\beta)^3}\right) \times \nonumber \\
&& \left(1-\frac{s_\sigma\gamma(1+3s_\beta z^4)}{(z^4-s_\beta)^3}\right) \ ,
\end{eqnarray}
with $G_z \equiv dG/dz$. Therefore, by formally integrating $G_{z}$, the metric component $g_{tt}$ in Eq.~(\ref{eq:lineel}) is obtained in terms of the radial function $z$ as
\begin{equation} \label{eq:gttz}
g_{tt}=-\frac{1}{f_R}\left(1-\frac{r_S (1+\delta_1 G(z))}{zr_{\star}f_R^{1/2}}\right) \ .
\end{equation}

\subsection{Geodesic completeness}\label{sec:geocom}

The non-trivial modified dynamics induced by the gravitational $R^2$ corrections necessarily modifies the geodesic structure of the corresponding geometry as compared to GR solution. This is a question of utmost interest, given the fact that geodesic completeness, namely, whether any (null and timelike) geodesic can be extended to arbitrarily large values of the affine parameter, is the most fundamental and generally accepted criterion to determine whether a spacetime is singular o not \cite{Wald}. Since timelike geodesics are associated to physical observers and null geodesics to the propagation of information, this criterion captures the intuitive idea that in a physically well behaved spacetime \emph{nothing can suddenly cease to exist} and that \emph{nothing can emerge out of nowhere}. Nonetheless, as discussed in the introduction, there is frequently a misunderstanding in the literature, taking curvature divergences as an equivalent concept to that of geodesic completeness in order to detect the presence of spacetime singularities. As we shall show in Sec.\ref{sec:V}, such an identification explicitly breaks in many of the geometries considered in this work. Thus we are mainly interested in studying the geodesic structure in those cases where the GR geodesics are incomplete and consequently yield a singularity, regardless of the presence or not of curvature divergences. To this end, in this section we shall specify the geodesic equation for Palatini $f(R)$ theories and solutions of the form studied here.

In a coordinate system, a geodesic curve $\gamma^{\mu}=x^{\mu}(\lambda)$ associated to a given connection $\Gamma^\mu_{\alpha\beta}$ is defined by the equation \cite{Wald}
\begin{equation}\label{eq:geodesics}
\frac{d^2x^\mu}{d\lambda^2}+\Gamma^\mu_{\alpha\beta}\frac{dx^\alpha}{d\lambda}\frac{dx^\beta}{d\lambda}=0 \ ,
\end{equation}
where $\lambda$ is the affine parameter. Since in the action (\ref{eq:action}) defining our model, the matter part couples to the metric but not to the connection, we will focus on the geodesics associated to the physical metric $g_{\mu\nu}$, which are the ones that the matter fields follow according to the Einstein equivalence principle (see \cite{OlmoBook} for an extended discussion on geodesics in metric-affine spaces).

The analysis can be largely simplified by writing the geodesic equation using the tangent vector $u^{\mu}=dx^{\mu}/d\lambda$, which satisfies $u_{\mu}u^{\mu}=k$, with $k=1,0,-1$ corresponding to spacelike, null, and timelike geodesics, respectively. Taking advantage of spherical symmetry, without loss of generality we can rotate the angular plane in such a way that it coincides with $\theta=\pi/2$, which further simplifies the problem. From the line element (\ref{eq:lineel}) we can, in addition, identify two conserved quantities of motion, $E=(A(x)/f_R) dt/d\lambda$ and $L=r^2(x) d\varphi/d\lambda$. For timelike geodesics, these quantities carry the meaning of the total energy per unit mass and angular momentum per unit mass, respectively. For null geodesics $E$ and $L$ lack a proper meaning by themselves, but the quantity $L/E$ can be identified as an apparent impact parameter as seen from the asymptotically flat infinity \cite{Chandra}.

Under these conditions, the geodesic equation (\ref{eq:geodesics}) for the above geometries simply reads

\begin{equation}\label{eq:geofR}
\frac{1}{f_R^2}\left(\frac{dx}{d\lambda}\right)^2=E^2-\frac{A(x)}{f_R}\left(\frac{L^2}{r^2(x)}-k\right) \ .
\end{equation}
By  using the relation of coordinates (\ref{eq:tildexz}) (and also the associated Eq.~(\ref{eq:dzdx})), we rewrite the geodesic equation (\ref{eq:geofR}) as
\begin{equation}
\frac{d\lambda}{dz} = \pm \frac{ f_R^{1/2}\left(1+\frac{zf_{R,z}}{2f_R}\right)}{\sqrt{E^2 f_R^2 - A(z)f_R
\left(\frac{L^2}{r_{\star}^2 z^2} - k\right)}} \ ,
\label{eq:geoz}
\end{equation}
where $\lambda$ is measured in units of $r_{\star}$, and the sign $\pm$ corresponds to outgoing/ingoing geodesics,  with
\begin{equation}
A(z)=1-\frac{r_S}{r_{\star}}\left(\frac{1+\delta_{1}G(z)}{z f_{R}^{1/2}}\right) \ ,
\label{eq:Az}
\end{equation}
as one can deduce by following the steps of Sec. \ref{subsec:sol}. Equivalently, the geodesic equation can be written in the more convenient form
\begin{equation}
\frac{d\lambda}{dz} = \pm \frac{\left(1+\frac{zf_{R,z}}{2f_R}\right)}{ f_R^{1/2}\sqrt{E^2 +g_{tt}
\left(\frac{L^2}{r_{\star}^2 z^2} - k\right)}} \ .
\label{eq:geozgtt}
\end{equation}

In the next section we shall study in detail the properties of the four different cases of configurations, corresponding to the combinations of the signs of $\sigma$ and $\beta$, and their respective features regarding the behaviour of the energy density, the curvature scalars, and geodesic completeness.

\section{Analysis of the solutions} \label{sec:V}

\subsection{Case I: $\sigma>0$, $\beta<0$ }\label{sec:pm}

\begin{figure}[h]
\begin{center}
\includegraphics[width=0.45\textwidth]{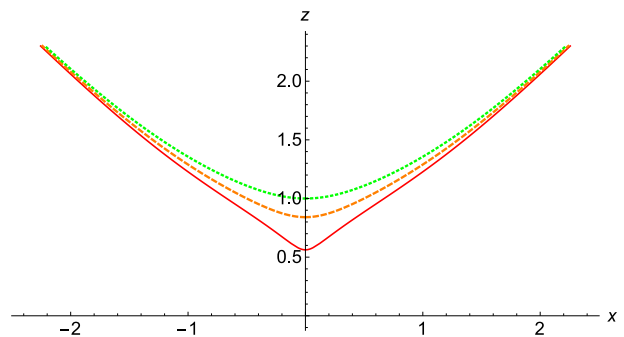}
\end{center}
\caption{Representation of $z(x)$ as a function of the radial coordinate $x$  (in units of $r_{\star}=\tilde{\beta}^{1/4}r_0$), for $\gamma=1.1$ (solid, red), $\gamma=1.5$ (dashed, orange) and $\gamma=2$ (dotted, green). Note that far from the bouncing region ($z=z_c$, $x=0$), where the wormhole throat is located, we have $z^2 \simeq x^2$, which restores the GR behavior there.}\label{fig:1}
\end{figure}

Let us now particularize the above equations to the case in which $s_\sigma=1$ and  $s_\beta=-1$, for which we obtain
\begin{eqnarray}
\rho&=&\frac{\rho_m}{z^4+1} \ ,\\
f_R&=& 1-\frac{\gamma }{(z^4+1)^2} \ , \label{eq:fzm}\\
G_z&=& \frac{z^2 \left(1-\frac{\gamma  \left(1-3 z^4\right)}{\left(z^4+1\right)^3}\right) \left(1-\frac{\gamma }{\left(z^4+1\right)^3}\right)}{\left(z^4+1\right) \left(1-\frac{\gamma }{\left(z^4+1\right)^2}\right)^{3/2}} \ . \label{eq:GzfRmp}
\end{eqnarray}
The function $G(z)$ determined by Eq.~(\ref{eq:GzfRmp}) can be easily solved using power series expansions, and the resulting solutions can be classified in terms of the values of the parameter $\gamma$ defined in Eq.~(\ref{eq:fz}). Depending on whether $\gamma$ is greater or smaller than unity, one finds different families of solutions. In this sense, the behavior of the function $z=z(\tilde{x})$, which arises from the resolution of Eq.~(\ref{eq:tildexz}), contains valuable information. Note that according to Eq.~(\ref{eq:fzm}) the function $f_{R}$ vanishes at
\begin{equation}  \label{eq:zc}
z_{c}=(\gamma^{1/2}-1)^{1/4} \ ,
\end{equation}
which sets a critical value for $\gamma=1$. When $\gamma\ge 1$, the radial function $z(x)$ has a minimum at $z_{c}$ where, according to Eq.~(\ref{eq:tildexz}), $\tilde{x}=0$. (From now on, we drop the tilde from $\tilde{x}$ to lighten the notation). Though a compact expression for $z=z(x)$ is not easy to find in general, a series expansion around $z=z_c$ yields the result
\begin{equation}\label{eq:xapprox}
|x| \approx \sqrt{ \frac{8z_c^5}{1+z_c^4}}(z-z_c)^{1/2}  +\mathcal{O}[(z-z_c)^{3/2}]\ .
\end{equation}
From this expression one finds that $z\approx z_c+x^2(1+z_c^4)/(8z_c^5)$, which shows that  for $x>0$ the area of the $2$-spheres decreases with decreasing $x$, but for $x<0$ increases with decreasing $x$, with a minimum at $x=0$ ($z=z_c$). This behavior is clearly seen in Fig.~\ref{fig:1} where Eq. (\ref{eq:tildexz}) has been inverted numerically for several values of $\gamma>1$. The interpretation of this minimal area in the two-spheres is well known in the literature: it represents a wormhole \cite{Visser}, a topologically non-trivial bridge connecting two asymptotically flat spacetime regions, where $z_c$ ($x=0$) sets the location of the throat. As it has been found in other cases of Palatini $f(R)$ theories coupled to various matter sources \cite{Universe,bcor16}, the emergence of this structure is directly related to the existence of zeros in the function $f_R$.

To study in more detail the geometry around $z_c$, it is useful to consider the following expansions
\begin{eqnarray}
f_R&\approx & \frac{8z_c^3}{1+z_c^4}(z-z_c) + \mathcal{O}[(z-z_c)^{2}] \ , \label{eq:fRg} \\
G_z&\approx & \frac{C}{(z-z_c)^{3/2}}  + \mathcal{O}[(z-z_c)^{-1/2}]\ , \label{eq:Gzg}
\end{eqnarray}
where $C=\frac{1}{\sqrt{32}}\frac{z_c^7}{[z_c(1+z_c^4)]^{3/2}}$ is a constant. Upon integration, one finds that
\begin{equation}
G(z)\approx \frac{-2C}{\sqrt{z-z_c}} + \mathcal{O}[(z-z_c)^{1/2}]\ ,
\label{eq:Gg}
\end{equation}
which diverges at $z=z_c$.  The $g_{tt}$ component of the physical metric appearing in Eq.~(\ref{eq:gttz}) can thus be approximated as
\begin{equation}
g_{tt} \approx -\frac{r_S}{r_{\star}}\frac{\delta_{1}}{64(z-z_c)^2}  + \mathcal{O}[(z-z_c)^{-3/2}]\ .
\label{eq:gttwh}
\end{equation}
Due to the divergence in this metric component, curvature scalars generically diverge near $z=z_c$. Obviously, this behavior is shared by all those models in which the function $f_R$ has a single pole at $z=z_c$. It should be noted, however, that curvature divergences are not synonyms with spacetime singularities, as mentioned in the introduction (see Sec.\ref{sec:geocom} below).

Now, let us analyze the case $0<\gamma<1$ for which $z_c$ in (\ref{eq:zc}) has no real solutions. In this case,  $z$ belongs to the range $(0, +\infty)$ and no wormhole geometries are found. Near the center, $z=0$, the relation (\ref{eq:tildexz}) can be expanded as
\begin{equation}
x \approx     \sqrt{1-\gamma} \, z +\mathcal{O}(z^5)\ ,
\label{eq:xnowh}
\end{equation}
while $f_R$ becomes there
\begin{equation}
f_{R} \underset{z \to 0}{\approx}1-\gamma +\mathcal{O}(z^{4})\ ,
\label{eq:fRnowh}
\end{equation}
and the function $G_z$ is finite
\begin{equation}
G_{z} \underset{z \to 0}{\approx} \sqrt{1-\gamma}z^2+\mathcal{O}(z^{6}) \ .
\label{eq:Gznowh}
\end{equation}
In this case,  near the origin $G(z)$ can be approximated as
\begin{equation}
G(z) \underset{z \to 0}{\approx} -\frac{1}{\delta_c^{(\gamma)}}+\sqrt{1-\gamma}\frac{z^3}{3} +\mathcal{O}(z^{7}) \ ,
\label{eq:Gnowh}
\end{equation}
where $\delta_c^{(\gamma)}$ is a constant (different for each value of $\gamma$) whose value guarantees that the Reissner-Nordstr\"om solution of GR is recovered  in the far limit, $z \gg 1$.  The expansion of the metric component $g_{tt}$ around the center is
\begin{equation} \label{eq:gttg}
g_{tt} \underset{z \to 0}{\approx} - \frac{1}{(1-\gamma)} +\frac{r_S}{r_{\star}} \frac{(1-\delta_1/\delta^{\gamma}_c)}{(1-\gamma)^{3/2}}\frac{1}{z}+\mathcal{O}(z^{2}) \ .
\end{equation}
Note that for the choice  $\delta_1=\delta_c^{(\gamma)}$, the metric is finite everywhere. On the other hand, for $\delta_1>\delta_c^{\gamma}$ the metric at the center is divergent and timelike, while for  $\delta_1<\delta_c^{\gamma}$ it becomes spacelike. Nonetheless, no matter the behaviour of the metric at the center, in all cases curvature invariants such as $K \equiv {R_\alpha}^{\beta\mu\nu}{R^\alpha}_{\beta\mu\nu}$ do always have divergences at $z=0$. The behaviour of the metric at the center also determines the number (and type) of the horizons, mimicking the basic description of some models of nonlinear electrodynamics \cite{BI-grav, NED-grav, dr}: two, one (degenerate) or no horizons for $\delta_1>\delta_c^{\gamma}$; a single non-degenerate horizon if $\delta_1<\delta_c^{\gamma}$, and no horizons if $\delta_1=\delta_c^{(\gamma)}$.

Finally, the critical case $\gamma=1$ must be treated separately, leading to
\begin{eqnarray}
f_{R} &\approx& 2z^4 -3z^8 +\mathcal{O}(z^{12}) \label{eq:fRgamma1}\\
G_{z} &\approx& \frac{9}{\sqrt{2}}z^4-\frac{117}{4 \sqrt{2}}z^8+\mathcal{O}(z^{12}) \ .
\label{eq:Gzgamma1} \\
G(z) &\approx& -\frac{1}{\delta_c^{(1)}}+\frac{9}{5 \sqrt{2}}z^5-\frac{13}{4 \sqrt{2}}z^9+\mathcal{O}(z^{12}) \ .
\label{eq:Ggamma1} \\
g_{tt}&\approx &-\frac{{r_S}}{r_{\star}}\frac{(\delta_1-\delta_c^{(1)})}{2\sqrt{2}\delta_c^{(1)} z^{7}}-\frac{1}{2z^{4}}+O(z^{-3}) \ . \label{eq:metricg1}
\end{eqnarray}
Thus the metric diverges at $z=0$, which induces the presence of curvature divergences there. Whether a wormhole exists in this case or not is a matter of taste, as its throat would have vanishing area:
\begin{equation}
x \approx    \sqrt{2}z^{3} -\frac{3}{\sqrt{8}}z^{7}+\mathcal{O}(z^{11})\ .
\label{eq:xgamma1}
\end{equation}

To summarize the results obtained so far, we can say that when the matter density scale $\rho_m$ is larger than the gravity scale $\rho_\sigma$, i.e., $\gamma=\rho_m/\rho_\sigma>1$, the theory yields wormhole solutions. Whether this wormhole is hidden behind an event horizon or not depends on the combination of parameters $\gamma, \delta_1, r_S$ characterizing the solutions. However, a detailed analysis of the horizon structure of these solutions is beyond the purpose of the present work. We just mention that the geometry is almost identical to the Reissner-Nordstr\"{o}m solution of GR everywhere except in the region within the inner horizon\footnote{Note, in this sense, that  in the asymptotic limit  $z \gg 1$, and for arbitrary $\gamma$, we have $f_R \simeq 1$ , so that $z^2 \approx \tilde{x}^2$ (we explicitly reintroduce the tilde here) and the role of $z$ as the radial coordinate in GR is restored, while the function $G(z)$ in Eq.~(\ref{eq:GzfR}) quickly converges to the GR solution, $G_z\approx 1/z^2$, thus recovering the Reissner-Nordstr\"om geometry of GR.}, where some departures arise and modify the structure of horizons. When the matter density scale $\rho_m$ is lower than the gravity scale $\rho_\sigma$, the wormhole throat closes $\gamma=1$ and no wormhole solution exist anymore $\gamma<1$.

\subsubsection*{Geodesic structure}\label{sec:geostruc}

Let us begin by noting that regardless of the value of $\gamma$, far from the center ($z \to +\infty$), $f_{R} \approx 1$ and $G_{z} \approx 1/z^{2}$. This means that in that region the GR solution is recovered and the geodesics are essentially coincident with those of GR there. One can verify numerically that this approximation is valid (almost exact!) for all configurations with $r_S/r_{\star}\ge 10$ and arbitrary $\delta_1$.

\begin{figure}[h]
\begin{center}
\includegraphics[width=0.45\textwidth]{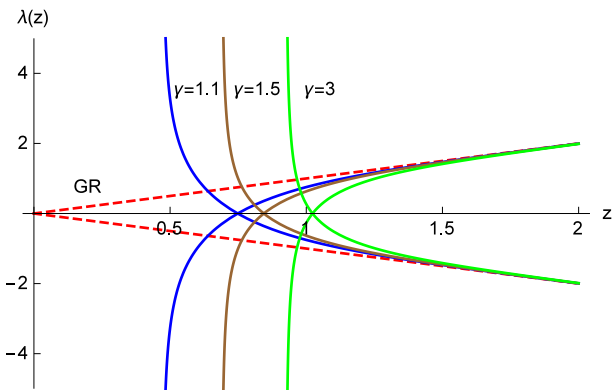}
\end{center}
\caption{Affine parameter $\lambda(z)$ (measured in units of $r_{\star}$) for radial null geodesics (with $E=1$), in the GR case ($\gamma=0$, dashed) and three $\gamma>1$ cases ($=1.1,1.5,3$, solid), for ingoing/outgoing trajectories. Far from $z=z_c(\gamma)$ (where the wormhole throat is located) all curves converge to the GR solution $\lambda(z)= z$, while for $z=z_c(\gamma)$, the curves $\lambda(z)$ (for $\gamma>1$) diverge to $\pm \infty$ which means that they are complete, as opposed to the GR case.
}
\label{fig:2}
\end{figure}

Let us focus first on the wormhole configurations, $\gamma>1$, for which our main concern is to study the deviations in the behavior of geodesics near the throat, located at  $z=z_c$. Consider first radial null geodesics ($k=L=0$). Near the wormhole throat $z \to z_c$ the geodesic equation (\ref{eq:geozgtt}) becomes
\begin{equation}\label{eq:geofR3}
\frac{d\lambda}{dz} \approx \pm \frac{\hat{C}}{(z-z_c)^{3/2}} \ ,
\end{equation}
where $\hat{C}=\sqrt{1+z_{c}^{4}}/(4E\sqrt{2z_c})$. By direct integration, we find
\begin{equation}
\lambda(z) \approx \mp \frac{2\hat{C}}{(z-z_c)^{1/2}} \ .
\end{equation}
From this expression it follows that as $z \to z_c$ one has $\lambda \to \pm \infty$. Stated in words, this means that ingoing light rays, emitted from $z \to +\infty$ when $\lambda \to -\infty$, approach the wormhole at $z \to z_c$  as $\lambda \to +\infty$, while outgoing light rays, which  propagate to $z \to +\infty$ as $\lambda \to + \infty$, set off from the wormhole at $z \to z_c$ and $\lambda \to -\infty$. A complete representation of the radial null geodesics is shown in Fig.~\ref{fig:2}.  From this plot one verifies that the far limit recovers the GR behavior while near the throat the affine parameter diverges, guaranteeing  in this way the completeness of these geodesics, in agreement with the analysis of the asymptotic behaviors provided above.

Let us now consider nonradial geodesics ($L\neq 0$) and/or timelike geodesics ($k=-1$). Since the left-hand side of (\ref{eq:geofR}) is positive by construction, physical trajectories must preserve the positivity of the right-hand side.  From the expansions (\ref{eq:fRg}) and (\ref{eq:Gg}) it follows that
\begin{equation}
A(z) \approx
\frac{r_S}{r_{\star}}\left(\frac{z_{c}^{3}}{8(1+z_{c}^{4})}\right)\frac{\delta_1}{z-z_c}
+ \mathcal{O}[(z-z_c)^{-1/2}] \ ,
\end{equation}
which diverges to $+\infty$ as $z\to z_c$. As a result, the right-hand side must vanish at some $z>z_c$, forcing in this way the bounce of these curves and  preventing them from reaching the wormhole throat. This is analogous to the behavior observed in the Reissner-Nordstr\"om solution of GR, where all such geodesics meet an infinite potential barrier generated by the central object \cite{Chandra} and never reach the central singularity. We thus conclude that all null and timelike geodesics are complete in this wormhole spacetime\footnote{One can check that spacelike geodesics are also complete in these wormhole spacetimes. Some of them never reach the wormhole but others can go through it. The latter correspond to $E=0=L$.}.

Let us recall that at the wormhole throat, $z=z_c$, curvature divergences arise. However, due to the fact that radial null geodesics take an infinite affine time to reach the throat, this implies that they lie at the boundary of the spacetime and do not belong to the physically accessible region. This way such divergences have no influence upon physical observers and there is no need to invoke any cosmic censorship conjecture or similar arguments to hide such configurations behind an event horizon. Since the wormhole throat cannot be causally reached in finite affine time, these results put forward the existence of explicit examples where the presence of curvature divergences do not unavoidably entail singular solutions.

Let us now consider those cases with $0< \gamma <1$ for which no wormhole structure was found. From Eqs. (\ref{eq:fRnowh}) and (\ref{eq:Gnowh}), we obtain for $z \to 0$ that
\begin{equation}
A(z)\approx -\frac{r_S}{r_{\star}}\left(\frac{1-\delta_1/\delta_c^{(\gamma)}}{\sqrt{1-\gamma}}\right)\frac{1}{z} + 1+\mathcal{O}(z^{2}) \ .
\end{equation}
The full discussion of the geodesic structure would proceed now in much the same way as in the case of certain models of non-linear electrodynamics coupled to GR, where the nature of the central region (spacelike or timelike), which depends on the ratio $\delta_1/\delta_c^{(\gamma)}$, will determine the type of geodesic able to approach the innermost region. Nonetheless, it is enough to consider radial null geodesics, for which the geodesic equation (\ref{eq:geoz}) reads
\begin{equation}
\frac{d\lambda}{dz} \approx \pm \frac{1}{E (1-\gamma)^{1/2}}\ .
\end{equation}
This equation can be readily integrated, $\lambda(z)=\lambda_0\pm \frac{z}{E (1-\gamma)^{1/2}}$, implying that the origin can be reached in a finite affine time, without possibility of further extension. This result is identical to that found in the GR case, which is regarded as singular, but is in sharp contrast with the previous results for the wormhole case.

\begin{figure}[h]
\begin{center}
\includegraphics[width=0.45\textwidth]{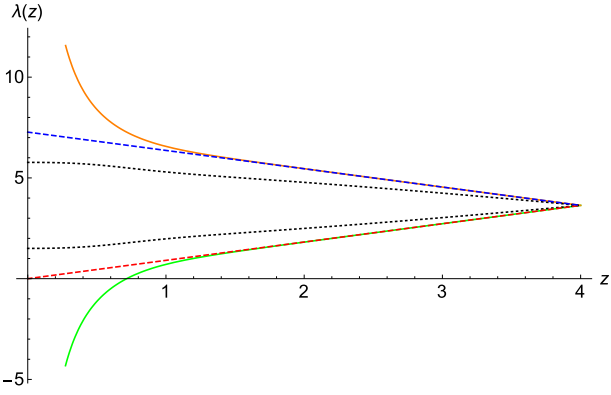}
\end{center}
\caption{Representation of the affine parameter $\lambda(z)$ as a function of the radial coordinate $z$  (in units of $r_{\star}=\tilde{\beta}^{1/4}r_0$), for $\gamma=1$. The dashed curves represent radial null geodesics in GR, while the solid ones are those of our gravity model. The upper/lower curve is the ingoing/outgoing light ray. Radial timelike geodesics with $\delta_1< \delta_c^{(1)}$ (black dotted curves) lie within the light cone and hit the origin in a finite affine time.}\label{fig:geodesicsgamma1}
\end{figure}

Finally, for the transition case $\gamma = 1$ (with $z_c=0$), the expansion of the metric as $z \to 0$ yields
\begin{equation}
A(z) \approx - \frac{r_{S}}{r_{\star}}\left(\frac{1-\delta_1/\delta_{c}^{(1)}}{\sqrt{2}}\right)\frac{1}{z^{3}} +1+ \mathcal{O}(z) \ .
\end{equation}
Radial null geodesics satisfy the following equation
\begin{equation}
\frac{d\lambda}{dz} \approx \pm \frac{r_{\star}}{\sqrt{2} E z^{2}}\ ,
\end{equation}
and one can easily find that $\lambda(z) \approx \mp \frac{r_{\star}}{E}\frac{1}{\sqrt{2}z} + \lambda_{0}$, which puts forward that these geodesics are complete, as they take an infinite affine time to reach the center. However, this model hides an unusual complexity (see Fig.~\ref{fig:geodesicsgamma1}). Indeed, if one considers nonradial and/or timelike geodesics, configurations with $\delta_1\ge\delta_c^{(1)}$ lead to a bounce at some $z>0$, while for those with $\delta_1< \delta_c^{(1)}$ geodesics take a finite affine time to reach the origin. Thus, despite the completeness of radial null geodesics and of those with $\delta_1\ge\delta_c^{(1)}$, the case $\gamma=1$ may lead to geodesically incomplete configurations, depending on the values of the parameters.

\subsection{Case II: $\sigma>0$, $\beta>0$ }\label{sec:pp}

\begin{figure}[h]
\begin{center}
\includegraphics[width=0.45\textwidth]{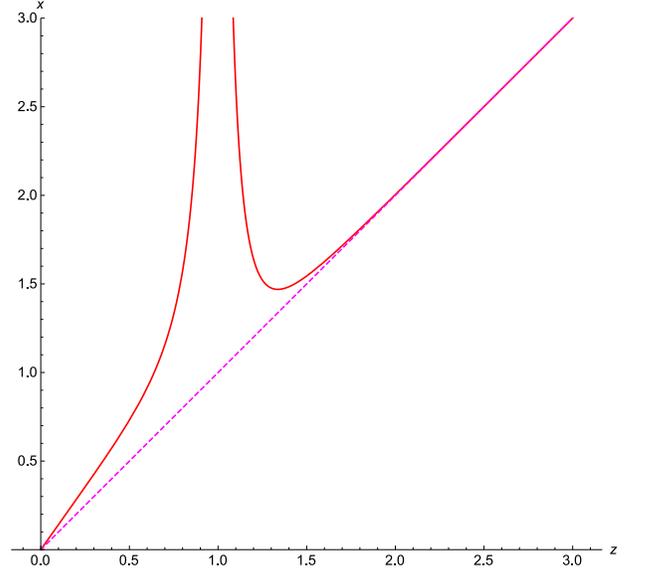}
\end{center}
\caption{Representation of $x(z)$ as a function of the radial coordinate $z$  (in units of $r_{\star}=\tilde{\beta}^{1/4}r_0$), for $\gamma=1$ (solid, red). The curve $x=z$ (dashed) is shown for comparison. We identify two regions where $x(z)$ is monotonic: one which tends to the GR solution $x=z$ and has a minimum, and another one close to the origin representing a non asymptotically flat solution which may be disconnected from the exterior one.}\label{fig:minimum}
\end{figure}

\begin{figure}[h]
\begin{center}
\includegraphics[width=0.45\textwidth]{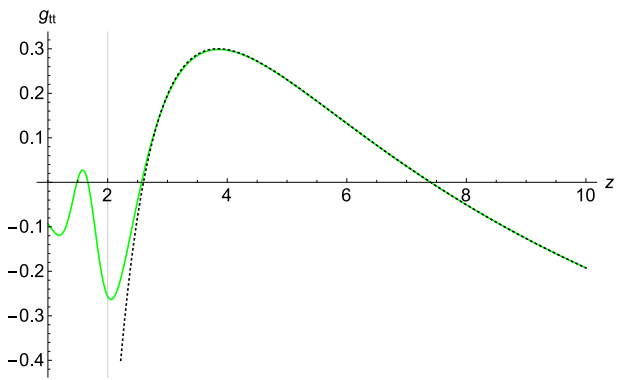}
\end{center}
\caption{Representation of $g_{tt}$ in the $\{\sigma>0,\beta>0\}$ case, as a function of the radial coordinate $z$, for $r_S/r_{\star}=10$ , $\delta_1=1.3(r_S/4r_{\star})$ and $z_{min}=2$ (solid curve). The dotted line represents the GR solution with the same parameters. Note that $g_{tt}$ cuts 4 times the $z-$axis, which implies $4$ horizons. The two external horizons are almost coincident with the GR prediction. The other two are a result of the modified dynamics. The vertical grey line represents the location where $x(z)$ reaches its minimum, which is (close but) unrelated to the minima of $g_{tt}$. Note that the horizontal axis begins at $z=1$.}\label{fig:metricSp}
\end{figure}

Let us now shift our attention to the case in which both $s_\sigma$ and $s_\beta$ are positive. Then we find
\begin{eqnarray}
\rho&=&\frac{\rho_m}{z^4-1} \\
f_R&=& 1+\frac{\gamma }{(z^4-1)^2} \label{eq:fzm}\\
Gz&=& \frac{z^2 \left(1-\frac{\gamma  \left(3 z^4+1\right)}{\left(z^4-1\right)^3}\right) \left(1-\frac{\gamma }{\left(z^4-1\right)^3}\right)}{\left(z^4-1\right) \left(\frac{\gamma }{\left(z^4-1\right)^2}+1\right)^{3/2}} \ . \label{eq:GzfRm}
\end{eqnarray}
An important difference as compared to the previous case is that, regardless of the value of $\gamma$, $f_R$ and $G_z$ diverge as $z\to 1$, where the energy density $\rho$ becomes infinite. Whether these divergences imply that the spacetime is singular or not is something nontrivial which must be determined after a careful scrutiny of the geometry and its geodesic structure. But before getting into that, one should note that  the relation between the coordinates $x$ and $z$, determined by $x^2=z^2 f_R$, now is not monotonic, having a minimum as shown in Fig.~\ref{fig:minimum}.
Unlike in the $\beta<0$ case, now the minimum is in the function $x=x(z)$ rather than in $z(x)$. This puts forward that now it is the auxiliary geometry which is of the wormhole type. This minimum represents the throat of the wormhole and its location is given by $x_{min}=2z_{min}^3/\sqrt{1+3z_{min}^4}$, where $z_{min}$ is related to $\gamma$ via $\gamma=(z_{min}^4-1)^3/(1+3z_{min}^4)$.  It should be noted that this wormhole is not symmetric, having an asymptotically Minkowskian region as $x\approx z \to \infty$ and a non flat region as $x\to \infty$ when $z\to 1$.
In fact, as $z\to 1$ in the internal region, $G_z\approx \gamma^{\frac{1}{2}}/64(z-1)^4$ and $f_R\approx \gamma/16(z-1)^2$ imply that $G(z)\approx -\gamma^{\frac{1}{2}}/[192(z-1)^3]$, which leads to
\begin{equation}
A(z)\approx \frac{r_S}{48r_{\star}}\frac{\delta_1}{(z-1)^2} \ .
\end{equation}
Using this relation and noting that as $z\to 1$ we have $x^2\approx \gamma/16(z-1)^2\to \infty$, one gets
\begin{equation}
h_{tt}\approx  - \frac{r_S}{r_{\star}}\frac{x^2}{4\gamma} \ ,
\end{equation}
which is timelike and divergent as $x\to \infty$ ($z\to 1$). The physical metric, on the other hand, has a completely different behavior. Given that $g_{tt}=h_{tt}/f_R$, and that $g_{rr}dr^2=g_{xx}dx^2$, expanding about $z\to 1$ yields
\begin{eqnarray}
g_{tt}&\approx&  - \frac{r_S\delta_1}{r_{\star}3\gamma}-\frac{2r_S(z-1)}{r_{\star}3\gamma}+O(z-1)^2 \ , \\
g_{rr}&\approx &  \frac{48r_{\star}}{r_S\delta_1}+O(z-1)^2\ ,
\end{eqnarray}
which are always finite at $z=1$ where, recall, the energy density diverges. In Fig.~\ref{fig:metricSp} the behavior of the $g_{tt}$ component is shown for a configuration which exhibits up to four horizons in the $z>1$ interval.

\begin{figure}[h]
\begin{center}
\includegraphics[width=0.45\textwidth]{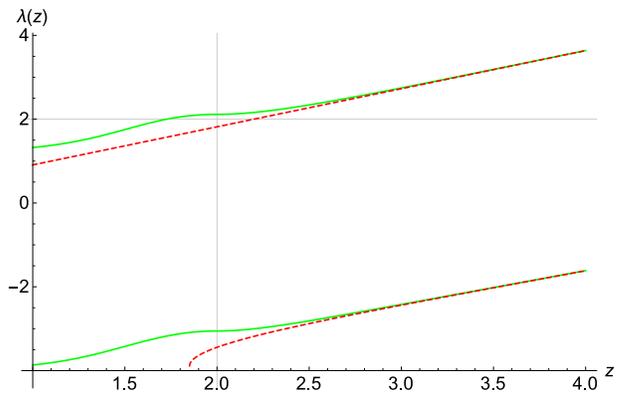}
\end{center}
\caption{Representation of the affine parameter $\lambda(z)$ in the $\{\sigma>0,\beta>0\}$ case, corresponding to radially ($L=0$) infalling geodesics with  $E=1.1$ in the geometry of Fig. \ref{fig:metricSp} (solid curves) as compared to their GR counterparts (dashed curves). The upper pair represents null geodesics while the lower one is timelike. Deviations from GR (dashed curve) only arise near $z_{min}$. In GR timelike observers bounce before reaching the center, while light rays get there in a finite affine time. In the quadratic $f(R)$ theory, both null and timelike geodesics reach the surface $z=1$ in a finite affine time. The vertical grid line represents the location where $x(z)$ reaches its minimum, which sets a saddle point for $\lambda(z)$. Note that the horizontal axis begins at $z=1$.}\label{fig:AffineSp}
\end{figure}

By proceeding in the same way as in the previous case, we now perform the analysis of the geodesic structure. In this sense, Fig. \ref{fig:AffineSp} puts forward that a region of infinite energy density is reached by null and timelike radial geodesics in a finite affine time. If the divergence in the matter sector is interpreted as defining a limiting boundary of the physical spacetime, where the equations no longer make sense, then the fact that geodesics can reach it in a finite affine time would imply that this geometry is singular.

From the numerical results shown in Figs. \ref{fig:metricSp} and \ref{fig:AffineSp} and, from the above analytical approximations, it is evident that nothing special happens to the physical metric at the points $z=1$ or $z=z_{min}$. This can be further emphasized by looking at the whole line element in the $z\to 1$ region, whose form is
\begin{equation}
ds^2\approx- \frac{r_S\delta_1}{r_{\star}3\gamma}dt^2+ \frac{48r_{\star}}{r_S\delta_1}dr^2+r_\star^2 d\Omega^2 \ .
\end{equation}
Using this line element, one readily verifies that all curvature invariants are finite at $z=1$ despite the energy density being divergent at that point. Though this divergence in the matter sector must be seen as a breakdown in the description of the fluid model considered, it serves to illustrate that divergences in the matter sector do not necessarily imply divergent curvature invariants. At the same time, the finiteness of curvature invariants is unrelated to the completeness of geodesics.

\subsubsection*{Analytical extension to $z<1$}\label{sec:extension}

The fact that the energy density diverges at $z=1$ and that it changes sign in the $z<1$ region somehow suggests that the physical region should be restricted to the open interval $z> 1$. Given that both null and timelike radial geodesics reach the $z=1$ surface in a finite affine time, if one wants to have a nonsingular spacetime, an artificial wormhole extension attached at $z=1$ should be considered to complete the geodesics. This construction, though mathematically admissible, seems a bit unnatural as compared to the wormhole solutions found in the $\{\sigma>0,\beta<0\}$ case, where the energy density is always finite.

\begin{figure}[h]
\begin{center}
\includegraphics[width=0.45\textwidth]{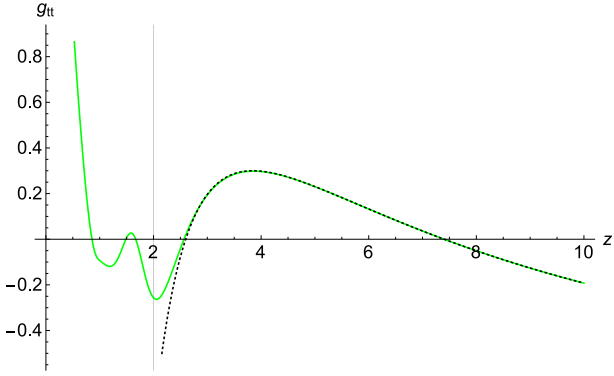}
\end{center}
\caption{Representation of $g_{tt}$ extended into the $z<1$ region for the same parameters as in Fig.~\ref{fig:metricSp}. The geometry now exhibits up to $5$ horizons. The metric diverges as $1/z$ as $z\to 0$ (Schwarzschild like). Note that the energy density in the region $z<1$ is always negative, with a divergence at $z=1$. In our view, only the region $z>1$ should be regarded as physical.}\label{fig:metricSp5h}
\end{figure}

\begin{figure}[h]
\begin{center}
\includegraphics[width=0.45\textwidth]{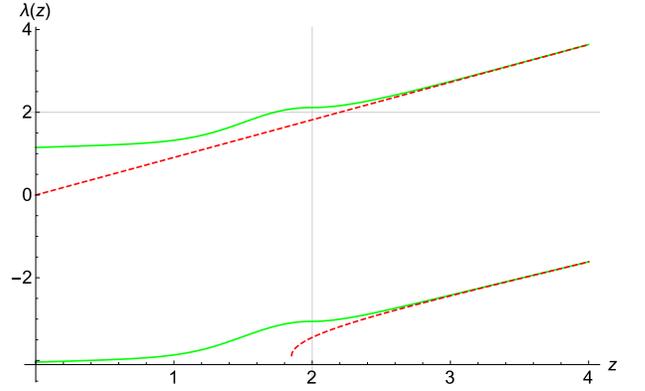}
\end{center}
\caption{Extension of the geodesics shown in Fig. \ref{fig:AffineSp} down to $z=0$. Both timelike and radial null geodesics take a finite affine time to reach the origin. Since the metric in this (unphysical) region diverges as $g_{tt}\sim 1/z$, the behavior is analogous to that found in the Schwarzschild black hole. }\label{fig:AffineSp5hor}
\end{figure}

Nonetheless, let us note that the numerical integration of the equations that define the geometry is well defined everywhere except at the point $z\to 0$. Moreover, the fact that $f_R$ and $G_z$ develop divergences at $z=1$ is not an obstacle to extend the integration across the $z=1$ surface, as a simple change of variables avoids the numerical difficulties associated to the parametrization in terms of $G(z)$. In fact, since $g_{tt}$ is well defined even at $z=1$, one can write a smooth differential equation for that function. The point is that the divergence in $G_z$ is somehow compensated by the divergence in $f_R$. Considering the ansatz (\ref{eq:gttz}) to isolate the function $G(z)$, one can compute its derivative and express it in terms of $g_{tt}$ and its first derivative. The resulting equation can be multiplied by $(z^4-1)^4$ on both sides to get rid of all the divergent terms. This new equation can be numerically integrated from the region where it coincides with GR down to $z=0$. The result is shown in Fig. \ref{fig:metricSp5h} and is in complete agreement with Fig. \ref{fig:metricSp} in the overlapping region $z\ge 1$. The corresponding extension of the geodesics appears in Fig. \ref{fig:AffineSp5hor}. Obviously, this region with $z<1$ does not make any physical sense, as it implies a negative energy density that changes from $+\infty$ on $z \gtrsim 1$ to $-\infty$ on $z \lesssim 1$ and remains negative until $z\to 0$. Nonetheless, the numerical problem is well defined all over the $z>0$ domain.

\subsection{Case III: $\sigma<0$, $\beta<0$ }\label{sec:mm}

\begin{figure}[h]
\begin{center}
\includegraphics[width=0.45\textwidth]{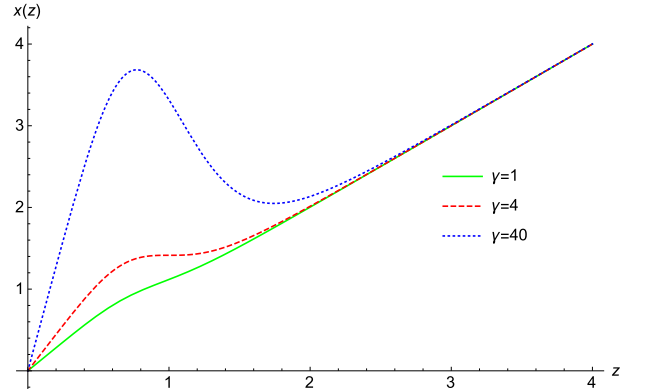}
\end{center}
\caption{Representation of $x(z)$ for different values of $\gamma$ in the model with $s_\sigma=-1=s_\beta$. }\label{fig:XdeZmm}
\end{figure}

When $s_\sigma=-1$ and $s_\beta=-1$, the model is characterized by
\begin{eqnarray}
\rho&=&\frac{\rho_m}{z^4+1} \ , \\
f_R&=& 1+\frac{\gamma }{(z^4+1)^2} \ ,\label{eq:fzm3}\\
G_z&=& \frac{z^2 \left(1+\frac{\gamma  \left(1-3 z^4\right)}{\left(z^4+1\right)^3}\right) \left(1+\frac{\gamma }{\left(z^4+1\right)^3}\right)}{\left(z^4+1\right) \left(1+\frac{\gamma }{\left(z^4+1\right)^2}\right)^{3/2}} \ . \label{eq:GzfRm3}
\end{eqnarray}
As follows from these expressions, the energy density and the functions $f_R$ and $G_z$ are everywhere smooth and finite. One can verify that for $\gamma>4$ the function $x(z)$ has a minimum at $z>1$ given by $\gamma=\frac{(1+z_{min}^4)^3}{3z_{min}^4-1}$. For this critical value of $\gamma$, the function $G_z$ has a zero at $z_{min}$, while for $\gamma>4$ it has two. The former occurs at $z_{min}$, while the latter appears at $0<z_{max}\leq 1\leq z_{min}$ and represents a local maximum (see Fig. \ref{fig:XdeZmm}).

Unlike in the case of Sec.\ref{sec:pp}, the minimum in $x(z)$ cannot be regarded as representing a wormhole in the auxiliary metric because it is not an absolute minimum in the sense that $x$ does not bounce back to infinity after crossing $z_{min}$. In the current case, below $z_{min}$  (see Fig. \ref{fig:XdeZmm}) $x(z)$ grows just until it reaches a new local extremum, now a maximum, after which it goes monotonically towards $x\to 0$ as $z\to 0$. The interpretation of this structure in the auxiliary geometry is unclear and will not be explored further in this paper. Since the existence of an internal wormhole or other structures in the auxiliary geometry does not impose any restriction on the radial function $z$ that defines the area of the two-spheres in the physical spacetime, we will assume that $z$ is naturally defined over the whole region $z\ge 0$.

An expansion of the metric in the $z\to 0$ region leads to

\begin{eqnarray}
g_{tt}&\approx& -r_S\frac{(\delta_1/\delta_c^{(\gamma)}-1)}{(1+\gamma)^{\frac{3}{2}}}\frac{1}{z}-\frac{1}{1+\gamma}\left(1-\frac{r_S}{r_{\star}}\frac{\delta_c^{(\gamma)}}{3}z^2\right)  \\
g_{rr}&\approx& \sqrt{1+\gamma}\frac{r_{\star}}{r_{S}}\left[\frac{z}{(\delta_1/\delta_c^{(\gamma)}-1)} - \frac{r_{\star}}{r_S}\frac{z^2}{(\delta_1/\delta_c^{(\gamma)}-1)^2}\right] \ ,
\end{eqnarray}
which shows that $g_{tt}$ diverges at the center if $\delta_1\neq \delta_c^{(\gamma)}$. When $\delta_1=\delta_c^{(\gamma)}$, the above expansion is not valid and must be re-evaluated, leading to
\begin{eqnarray}
g_{tt}&\approx&\frac{1}{1+\gamma}\left(-1+\frac{1}{3}\frac{r_S}{r_{\star}}\delta_c^{(\gamma)}z^2\right) \ , \\
g_{rr}&\approx& 1+\frac{1}{3}\frac{r_S}{r_{\star}}\delta_c^{(\gamma)}z^2 \ ,
\end{eqnarray}
which is completely regular at $z=0$. In this particular case, one verifies that the geometry at $z=0$ becomes of de Sitter type, with $R_{\mu\nu}=\frac{\delta_c^{(\gamma)}}{\delta_2}g_{\mu\nu}$ and finite curvature scalars. It is also worth noting that by a simple rescaling of the time coordinate, $t\to \sqrt{1+\gamma}\tilde{t}$, the metric $g_{\mu\nu}$ becomes Minkowskian at the center, showing that these coordinates locally represent free falling observers.

By inspecting the geodesic equation (\ref{eq:geozgtt}) shows that if $\delta_1>\delta_c^{(\gamma)}$ then non-radial and timelike geodesics bounce before reaching to the center (like in the usual Reissner-Nordstr\"{o}m black hole of GR). However, if  $\delta_1<\delta_c^{(\gamma)}$ (Schwarzschild like configuration)  nothing prevents those geodesics from getting there in a finite affine time. Radial null geodesics also reach the center regardless of the value of $\delta_1$. It should be noted that curvature divergences exist at $z=0$ as long as $\delta_1\neq \delta_c^{(\gamma)}$. But note that these divergences arise despite the fact that the energy density is finite everywhere, as in Case I of Sec. \ref{sec:pm}.

On the other hand, when $\delta_1=\delta_c^{(\gamma)}$, we have seen above that the geometry near the center is of de Sitter type and, therefore, the geodesics reaching there should not experience any pathological effect, being able to cross the center and continue their path\footnote{Note in this sense that replacing the point-like singularity by a de Sitter core is a standard strategy in looking for regular solutions in the context of GR, see e.g. \cite{AdScore,Ansoldi}.}. Configurations with $\delta_1=\delta_c^{\gamma}$ should thus be regarded as nonsingular. In this sense, the fact that radial null geodesics are insensitive to the value of $\delta_1$ suggests that they should always be able to go through the {\it apparently pathological} region $z=0$ even when $\delta_1\neq \delta_c^{(\gamma)}$. This view is further reinforced by the lack of correlation between the behavior of curvature scalars, divergent for $\delta_1\neq \delta_c^{(\gamma)}$, and the energy density, which is always finite. In view of all this, it is unclear whether one should regard any of these solutions as singular, as is typically assumed in the case of GR. Therefore, further analysis on the impact of curvature divergences on the transit of physical observers and on the scattering of waves in these spacetimes to get deeper into their singular/nonsingular character seems necessary and will be carried out elsewhere.

\subsection{Case IV: $\sigma<0$, $\beta>0$ }\label{sec:mm}

\begin{figure}[h]
\begin{center}
\includegraphics[width=0.45\textwidth]{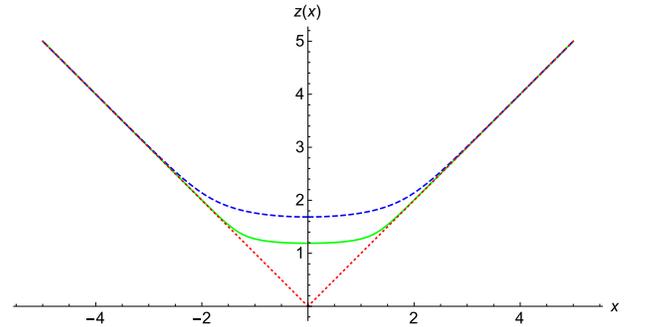}
\end{center}
\caption{Representation of $z(x)$ for $\gamma=1$ (solid green curve) and  $\gamma=50$ (dashed blue curve) as compared to the GR behaviour $z^2=x^2$ (dashed red line) in the model with $s_\sigma=-1$ and $s_\beta=1$. This shows that the physical metric describes a wormhole structure. Far from the central region the GR behavior is recovered $z^2(x) \approx x^2$.}\label{fig:WHmp}
\end{figure}

We now consider the last case, $s_\sigma=-1$ and $s_\beta=1$, which is characterized by
\begin{eqnarray}
\rho&=&\frac{\rho_m}{z^4-1} \ , \\
f_R&=& 1-\frac{\gamma }{(z^4-1)^2} \ , \label{eq:fzm4}\\
G_z&=& \frac{z^2 \left(1+\frac{\gamma  \left(1+3 z^4\right)}{\left(z^4+1\right)^3}\right) \left(1+\frac{\gamma }{\left(z^4-1\right)^3}\right)}{\left(z^4-1\right) \left(1-\frac{\gamma }{\left(z^4-1\right)^2}\right)^{3/2}} \ . \label{eq:GzfRm4}
\end{eqnarray}
In this model the energy density diverges at $z=1$ but, fortunately, this surface lies beyond the physically accessible region. Therefore, it can be considered finite beyond $z > 1$. Indeed, a glance at the function $f_R$ above puts forward that $z(x)$ has a minimum of magnitude $z_c=(1+\gamma^{1/2})^{1/4}$ at $x=0$, confirming in this way that this model describes wormholes (see Fig. \ref{fig:WHmp}) and that $z>1$ always.

A series expansion of the metric about $z=z_c$ leads to
\begin{equation}
g_{tt}\approx -\frac{r_S}{r_{\star}}\frac{\delta_1}{64}\frac{1}{(z-z_c)^2}+O(z-z_c)^{-3/2} \ ,
\end{equation}
which is always negative and divergent at the throat, implying the existence of curvature divergences on that surface. With this behavior near the throat, a glance to the geodesic equation (\ref{eq:geozgtt}) indicates that all timelike and non-radial geodesics ($L\neq 0$) bounce before reaching the wormhole, just like in the Reissner-Nordstr\"om black holes of GR. For radial null geodesics, however, one finds
\begin{equation}
E\frac{d\lambda}{dz}\approx \pm \frac{\sqrt{\frac{z_c^4-1}{z_c}}}{4 \sqrt{2} (z-z_c)^{3/2}} \ ,
\end{equation}
which implies that
\begin{equation}
E\lambda\approx \mp \frac{\sqrt{\frac{z_c^4-1}{z_c}}}{2 \sqrt{2} (z-z_c)^{1/2}} \ ,
\end{equation}
diverges as $z\to z_c$, confirming that such geodesics take an infinite affine time to reach (or come out from) the wormhole throat, similarly as in the case $\{\sigma>0,\beta<0\}$ of Sec.(\ref{sec:pm}). This model, therefore, always yields geodesically complete, nonsingular spacetimes. Note also that, though the geometry has curvature divergences at $z=z_c$, the energy density is finite there, taking the value $\rho(z_c)=\sqrt{\rho_m\rho_\sigma}$.

\section{Conclusions} \label{sec:VI}

\subsection{Overview of results}

In this work we have considered the problem of classical, non-rotating black holes in a quadratic Palatini $f(R)$ extension of GR. As matter fields we have taken an anisotropic fluid whose stress-energy tensor, besides satisfying the energy conditions, covers a number of interesting cases, in particular, that of non-linear models of electrodynamics. The latter have been frequently employed in the context of GR in order to take care of the singularity problem, but the strategy applied has been unsatisfactory so far. In contrast, the non-trivial interaction between gravity and matter in our case, which is encoded in the (gravity and matter) parameters $\sigma$ and $\beta$, yields a number of novelties on the corresponding spacetimes. These geometries quickly recover the Reissner-Nordstr\"om solution of GR far from the center, but drastically modify the innermost structure. The corresponding analysis is split into four different cases, according to the signs of $\sigma$ and $\beta$.

\begin{table*}
   \begin{tabular}{| c | c | c | c| c |c |}
        \hline
      &\footnotesize Wormhole &\footnotesize Metric components & \footnotesize Energy density & \footnotesize Curvature scalars & Geodesics \\
      \hline
      & & & & & \\
      \footnotesize Case I & \footnotesize YES if $\gamma>1$ & \footnotesize Divergent & \footnotesize Finite & \footnotesize Divergent & Complete if $\gamma>1$ \\
      ($\sigma>0,\beta<0$) & NO if $\gamma < 1$  & & & & Incomplete if $\gamma < 1$ \\
      & & & & & \\
      \hline
      & & & & & \\
      \footnotesize Case II & NO & \footnotesize Finite & Divergent & Finite & Incomplete \\
      ($\sigma>0,\beta>0$) & &  &  & & \\
      \hline
      & & & & & Incomplete? if $\delta_1>\delta_c^{(\gamma)}$ \\
      \footnotesize Case III & \footnotesize NO & \footnotesize Divergent if $\delta_1 \neq \delta_c^{(\gamma)}$ & \footnotesize Finite & \footnotesize Divergent if $\delta_1 \neq \delta_c^{(\gamma)}$ & Complete if $\delta_1=\delta_c^{(\gamma)}$ \\
      ($\sigma<0,\beta<0$) &  & Finite if $\delta_1 = \delta_c^{(\gamma)}$ & & Finite if $\delta_1 = \delta_c^{(\gamma)}$ & (de Sitter core) \\
      & & & & & Incomplete if $\delta_1<\delta_c^{(\gamma)}$ \\
      \hline
       & & & & & \\
      \footnotesize Case IV & YES & \footnotesize Divergent & Finite & Divergent & Complete \\
      ($\sigma<0,\beta> 0$) & &  &  & &  \\
      \hline
   \end{tabular}
   \caption{Summary of the features of the four families of configurations studied in Sec.\ref{sec:V} (the case $\{\sigma>0,\beta<0\}$ with $\gamma=1$ hides some peculiarities, see Sec. \ref{sec:pm}, so it is not contained in this table). The metric components, energy density and curvature scalars refers to the behaviour at the wormhole throat (when it exists) or otherwise at the innermost region of the solutions. Incomplete geodesics refer to the existence of (at least) a single incomplete null or timelike geodesic curve. The breakdown of the correlations among these three concepts is clear.}
 \label{table:I}
\end{table*}

\begin{itemize}
\item In the case $\{\sigma>0,\beta<0\}$ two classes of configurations are found, according to a certain critical scale, parameterized by $\gamma=1$. If $0<\gamma < 1$, the nature of the central point-like singularity (timelike, spacelike or null), is qualitatively similar to that found in certain families of models of non-linear electrodynamics, and the same applies to the structure of horizons \cite{dr}. In such a case one finds both divergence of curvature scalars and incompleteness of geodesics, hence these spacetimes are regarded as singular. But when $\gamma> 1$ the point-like singularity is shifted to a spherical surface of radius $z_c(\gamma)>0$, which corresponds to the minimum of the radial function $z(x)$ and represents the throat of a wormhole. Despite curvature scalars being divergent on this surface, we have explicitly shown that this does not prevent the completeness of geodesics for null and timelike observers, in such a way that the wormhole always lies on the future (or past) boundary or the manifold and cannot be reached in finite affine time by geodesic observers. The limiting case $\gamma=1$ has a number of peculiar features, such as complete radial null geodesics but incomplete non-radial and timelike geodesics for certain configurations (with $\delta_1< \delta_c^{(1)}$). This allows us to conclude that the cases with $\gamma\leq 1$ represent singular spacetimes, while those with $\gamma>1$ are nonsingular and possess a wormhole structure. Nonetheless, note that in all these cases (both nonsingular and singular) the energy density of the fluid is everywhere finite.

\item For the case $\{\sigma>0,\beta>0\}$ one finds that it is the auxiliary metric which has a wormhole at $z=1$, while the physical metric becomes finite there, exhibiting up to four horizons. The $z=1$ surface is a region of divergent energy density of the fluid, which is reached in finite affine time by null and timelike geodesics, but nonetheless the curvature scalars are finite there. The metric itself admits an analytical extension to the region $z<1$, but the spacetime would still be geodesically incomplete and, besides, the breakdown of the fluid model at $z=1$ seems to indicate the unphysical character of such an extension.

\item In the case $\{\sigma<0,\beta<0\}$ no wormhole solution is found and the metric is naturally extended down to $z=0$. An expansion of the physical metric there reveals that curvature divergences are present as long as $\delta_1 \neq \delta_c^{(\gamma)}$, but do disappear for the choice $\delta_1 = \delta_c^{(\gamma)}$, for which the metric at the center has a de Sitter behavior. Nonetheless, in both cases the energy density is everywhere finite. In the finite curvature cases, $\delta_1 = \delta_c^{(\gamma)}$, radial null geodesics reaching the $z=0$ region can clearly continue their path. Given that  these geodesics are insensitive to the value of $\delta_1$, one is tempted to interpret all such radial geodesics as complete, including the cases with $\delta_1\neq \delta_c^{(\gamma)}$. This interpretation is appealing but risky, as it would logically lead to the conclusion that the Reissner-Nordstr\"{o}m black hole of GR is nonsingular (as far as geodesic motion is concerned). To deepen into this question, an analysis on the impact of curvature divergences on congruences of observers and on the propagation of waves would be necessary along the lines of Refs.\cite{sw} and \cite{congruences}.

\item For the case $\{\sigma<0,\beta>0\}$, it always yields wormhole structures that naturally provide geodesically complete spacetimes, much in the same way as in the $\{\sigma>0,\beta<0\}$ case for $\gamma>1$. Note that this is so despite the fact that curvature divergences are always present at $z=z_c$, but the energy density is finite there.
\end{itemize}

In table \ref{table:I} we display the most relevant features of the four classes of configurations classified according to the combinations of signs of $\sigma$ and $\beta$. Such features are i) the existence or not of wormhole configurations and the behaviour at the wormhole throat (or at the center when no such wormhole exists) of the following objects: ii) the metric components , iii) the energy density, iv) the curvature scalars, and  v) the completeness (or not) of all null and timelike geodesics on such spacetimes. From this table it is clear the breakdown of the correlations between the behaviour of energy density, curvature scalars and completeness of geodesics in different ways.

\subsection{Final comments} \label{sec:VII}

The research presented in this work is added to the growing set of results within metric-affine geometries supported by $f(R)$ gravity \cite{Universe,bcor16}, quadratic $f(R,R_{\mu\nu}R^{\mu\nu})$ and Eddington-inspired Born-Infeld gravities \cite{sw,BIreg}, where the point-like singularity of GR is replaced by a wormhole structure, which allows geodesics to be complete. The results obtained so far seem to point out two different mechanisms for the resolution of spacetime singularities in this context. (i): For $f(R)$ theories the wormhole lies on the future (or past) boundary of the spacetime, but the fact that geodesics can be indefinitely extended means that these spacetimes are nonsingular according to the criteria employed in the singularity theorems \cite{Theorems}. ii): For the cases of quadratic and Born-Infeld gravity (which admit extensions to higher \cite{hd} and lower \cite{3D} dimensions with similar results), the wormhole may be reached in a finite affine time by some geodesics, but they are naturally extended through it. It should be stressed that in the latter case one may wonder about the physical impact of curvature divergences on physical (extended) bodies crossing the wormhole throat, since the presence of large tidal forces could rip them apart. In this sense, in four-dimensional Born-Infeld gravity a separate analysis using a congruence of geodesics to model such extended bodies has revealed that curvature divergences occurring at the wormhole throat do not pose any destructive threat, being the transit smooth and weakly affected by the large tidal forces at the throat \cite{congruences}. The problem of scalar wave scattering off this wormhole turns also to be well posed \cite{sw}.

To conclude, the results obtained in this paper provide a fauna of examples on which the (typically assumed) correlations among curvature divergences, energy density,  and geodesic (in)completeness are explicitly broken. The emergence of structures such as wormholes appears as a key element for the extendibility of geodesics (note that we also found a de Sitter interior which guarantees the extendibility of geodesics in a particular example). These non-perturbative features should be taken into account in  the interpretation of the existence of divergences in curvature scalars or in the matter fields. Those infinities are significant in perturbative frameworks, where they signal the end of validity of a certain approximation, but in a non-perturbative scenario their interpretation is unclear. It is worth noting, in this sense, that in systems such as graphene, wormholes have been constructed in the discrete description through a careful design of the lattice,  adding a series of heptagonal rings that help join two flat sheets with a carbon nanotube \cite{grapheneWH}. In the continuized description, the wormhole exhibits a curvature divergence at the throat which does not prevent the study of fermion propagation in the resulting effective geometry. This example is, in our view, very representative of what may correspond to a quantum version of the models presented here: geodesic completeness is guaranteed by non-perturbative aspects (the wormhole), though infinities (in curvature and energy density) may arise because of the lack of control of the underlying microstructure in a certain limit \cite{crystals}. The behavior of classical and quantum fields in these backgrounds is currently underway.

\section*{Acknowledgments}

C. B. is funded by the National Scientific and Technical Research Council (CONICET). G. J. O. is supported by a Ramon y Cajal contract, the i-COOPB20105 grant of the Spanish Research Council (CSIC), the Consolider Program CPANPHY-1205388, and the Severo Ochoa grant SEV-2014-0398 (Spain). D. R.-G. is funded by the Funda\c{c}\~ao para a Ci\^encia e a Tecnologia (FCT, Portugal) postdoctoral fellowship No.~SFRH/BPD/102958/2014 and the FCT research grant UID/FIS/04434/2013. This work is also supported by the Spanish grant FIS2014-57387-C3-1-P from MINECO and the CNPq (Brazilian agency) project No.~301137/2014-5. C. B. and D. R.-G. thank the Department of Physics of the University of Valencia for their hospitality during the elaboration of this work. This article is based upon work from COST Action CA15117, supported by COST (European Cooperation in Science and Technology).


\begin{thebibliography}{99}

\bibitem{Theorems}
R. Penrose, Phys. Rev. Lett. \textbf{14}, 57 (1965);
Gen. Rel. Grav. \textbf{34}, 1141 (2002);
S. W. Hawking, Phys. Rev. Lett. \textbf{17}, 444 (1966);
J.~M.~M.~Senovilla and D.~Garfinkle, Class.\ Quant.\ Grav.\  {\bf 32}, 124008 (2015).
\bibitem{CCC}
R. Penrose, Riv. Nuovo Cim. Numero Speciale \textbf{1}, 252 (1969).
\bibitem{Geroch}
R. Geroch, Ann. Phys. \textbf{48}, 526 (1968).
\bibitem{Wald}
R. M. Wald, \emph{General Relativity} (The University of Chicago Press, Chicago and London, 1984).
\bibitem{Hawking-Ellis} S. W. Hawking and G. F. R. Ellis, \emph{The Large Scale Structure of spacetime} (Cambridge University Press, Cambridge,
England, 1973).
\bibitem{phy}
E. Curiel and P. Bokulich, {\it Singularities and Black Holes}, The Stanford Encyclopedia of Philosophy (\href{http://plato.stanford.edu/archives/fall2012/entries/spacetime-singularities/}{Fall 2012 Edition}), Edward N. Zalta (ed.).
\bibitem{HS}
G.~T.~Horowitz and R.~C.~Myers,  Gen.\ Rel.\ Grav.\  {\bf 27},  915 (1995).
\bibitem{Ansoldi}
S. Ansoldi, arXiv:0802.0330[gr-qc].
\bibitem{BI}
M. Born and L. Infeld, Proc. Roy. Soc. London. A \textbf{144}, 425 (1934).
\bibitem{BI-grav}
A. Garcia, H. Salazar and J. F. Plebanski, Nuovo. Cim. \textbf{84}, 65 (1984);
M. Demianski, Found. Phys. \textbf{16}, 187 (1986);
G. W. Gibbons and D. A. Rasheed, Nucl. Phys. B \textbf{454}, 185 (1995).
\bibitem{NED-grav}
H. P. de Oliveira, Class. Quant. Grav. \textbf{11}, 1469 (1994);
H. Yajima and T. Tamaki, Phys. Rev. D \textbf{63}, 064007 (2001);
N. Breton, Phys. Rev. D \textbf{67}, 124004 (2003);
I.~Z.~Stefanov, S.~S.~Yazadjiev, and M.~D.~Todorov, Mod.\ Phys.\ Lett.\ A {\bf 22},  1217 (2007);
Phys.\ Rev.\ D {\bf 75}, 084036 (2007);
M. Hassaine and C. Martinez, Phys. Rev. D \textbf{75}, 027502 (2007);
Class. Quant. Grav. \textbf{25}, 195023 (2008);
H.~A.~Gonzalez, M.~Hassaine, and C.~Martinez,  Phys.\ Rev.\ D {\bf 80},  104008 (2009);
S.~Habib Mazharimousavi, M.~Halilsoy and O.~Gurtug,  Eur.\ Phys.\ J.\ C {\bf 74}, 2735 (2014);
A.~Sheykhi and S.~Hajkhalili, Phys.\ Rev.\ D {\bf 89}, 104019 (2014);
A.~Sheykhi and A.~Kazemi, Phys.\ Rev.\ D {\bf 90}, 044028 (2014).
\bibitem{AB}
E. Ay\'on-Beato and A. Garc\'ia, Phys. Rev. Lett. \textbf{80}, 5056 (1998);
Gen. Rel. Grav. \textbf{31},  629 (1999).
\bibitem{Bronnikov}
K.~A.~Bronnikov, Phys.\ Rev.\ Lett.\  {\bf 85},  4641 (2000).
\bibitem{Novello}
M.~Novello, S.~E.~Perez Bergliaffa, and J.~M.~Salim,  Class.\ Quant.\ Grav.\  {\bf 17},  3821 (2000).
\bibitem{GB}
D. Lovelock, J. Math. Phys. \textbf{12}, 498 (1971);
N. Deruelle and L. Farina-Busto, Phys. Rev. D \textbf{41}, 3696 (1990);
C. Charmousis, Lec. Notes Phys. \textbf{769}, 299 (2008);
C. Garraffo and G. Giribet, Mod. Phys. Lett. A \textbf{23}, 1801 (2008).
\bibitem{GBNED}
D. L. Wiltshire, Phys. Rev. D \textbf{38}, 2445 (1988);
M. Aiello, R. Ferraro, and G. Giribet, Phys. Rev. D \textbf{70}, 104014 (2004);
M. H. Dehghani, N. Alinejadi, and S. H. Hendi, Phys. Rev. D \textbf{77}, 104025 (2008);
H. Maeda, M. Hassaine, and C. Martinez, Phys. Rev. D \textbf{79}, 044012 (2009);
S. H. Hendi, Phys. Lett. B \textbf{677}, 123 (2009);
O. Miskovic and R. Olea, Phys. Rev. D \textbf{83}, 024011 (2011);
S.~H.~Hendi and R.~Naderi, Phys.\ Rev.\ D {\bf 91}, 024007 (2015);
D. Rubiera-Garcia, Phys. Rev. D \textbf{91}, 064065 (2015).
\bibitem{RegularBGR}
W.~Berej, J.~Matyjasek, D.~Tryniecki, and M.~Woronowicz, Gen.\ Rel.\ Grav.\  {\bf 38}, 885 (2006);
C.~Rovelli and F.~Vidotto,  Int.\ J.\ Mod.\ Phys.\ D {\bf 23},  1442026 (2014);
C.~Barcel\'o, R.~Carballo-Rubio and L.~J.~Garay, Universe {\bf 2}, 7 (2016);
M.~E.~Rodrigues, J.~C.~Fabris, E.~L.~B.~Junior, and G.~T.~Marques,  Eur.\ Phys.\ J.\ C {\bf 76}, 250 (2016);
M.~E.~Rodrigues, E.~L.~B.~Junior, G.~T.~Marques, and V.~T.~Zanchin,  Phys.\ Rev.\ D {\bf 94}, 024062 (2016).
\bibitem{SSP}
C. Kittel, \emph{Introduction to Solid State Physics (8th edition)} (Wiley, 2005);
J. D. Clayton, \emph{Nonlinear mechanics of crystals} (Springer, 2011).
\bibitem{PL}
E. Kr\"oner, Trends in Applications of Pure Mathematics to Mechanics Lecture Notes in Physics, \textbf{249}, 281 (1986);
K. Kondo, Proc. 2nd Japan Kat. Congr. of Appl. at Max-Planek-lnstitut fur Mctallforschung, Postfach 800665, D-7000 Stuttgart 80, BRD.
Mechanics, p.41 (1952);
B. A. Bilby, R. Bullough, and E. Smith, Proc. Roy. Soc. London, Ser. A \textbf{231}, 263 (1955);
F. Falk, J. Elast. \textbf{11}, 359 (1981);
M. O. Katanaev and I. V. Volovich, Ann. Phys. \textbf{216}, 1 (1992);
A.~Iorio,  Int.\ J.\ Mod.\ Phys.\ D {\bf 24}, 1530013 (2015).
\bibitem{ORtorsion}
G. J. Olmo and D. Rubiera-Garcia, Phys. Rev. D \textbf{88},  084030 (2013).
\bibitem{Pal}
G. J. Olmo, Int.\ J.\ Mod.\ Phys.\  D {\bf 20}, 413 (2011).
\bibitem{TEGR}
J.~W.~ Maluf, Ann. Phys. \textbf{525}, 339 (2013);
J.~G. Pereira, Teleparallelism: A New Insight into Gravity in \emph{Handbook of spacetime}, A. Ashtekar and V. Petcov (Eds.) (Springer, Berlin, 2014)
\bibitem{EC}
E. Cartan, C. R. Acad. Sci. Paris \textbf{174}, 593 (1922);
734 (1922);
F. W. Hehl, P. von der Heyde, G. D. Kerlick, and J. M. Nester, Rev. Mod. Phys. \textbf{48}, 393 (1976);
J.~E.~Daum and M.~Reuter,  Annals Phys.\  {\bf 334}, 351 (2013);
K.~Atazadeh, JCAP {\bf 1406}, 020 (2014).
\bibitem{Zanelli}
J. Zanelli, arXiv:hep-th/0502193.
\bibitem{fRlit}
A. De Felice and S. Tsujikawa, Liv. Rev. Rel. \textbf{13}, 3 (2010);
S. Capozziello and M. De Laurentis, Phys. Rep. \textbf{509}, 167 (2011);
S. Nojiri and S. D. Odintsov, Phys. Rep. \textbf{505}, 59 (2011);
S.~Nojiri, S.~D.~Odintsov, and D.~Saez-Gomez, Phys.\ Lett.\ B {\bf 681}, 74 (2009);
A.~de la Cruz-Dombriz, and D.~Saez-Gomez, Entropy {\bf 14},  1717 (2012).
\bibitem{LIGO}
B.~P.~Abbott {\it et al.} [LIGO Scientific and Virgo Collaborations], Phys.\ Rev.\ X {\bf 6}, 041015 (2016).
\bibitem{ECOs}
V.~Cardoso, S.~Hopper, C.~F.~B.~Macedo, C.~Palenzuela, and P.~Pani,  Phys.\ Rev.\ D {\bf 94},  084031 (2016);
J.~Abedi, H.~Dykaar, and N.~Afshordi, arXiv:1612.00266 [gr-qc];
C.~Barcel\'o, R.~Carballo-Rubio, and L.~J.~Garay, arXiv:1701.09156 [gr-qc].
\bibitem{af}
V. V. Kiselev, Class. Quant. Grav. \textbf{20}, 1187 (2003);
W.~Florkowski, Phys.\ Lett.\ B {\bf 668}, 32 (2008);
M.~Sharif and H.~R.~Kausar, Phys.\ Lett.\ B {\bf 697}, 1 (2011);
M.~H.~Daouda, M.~E.~Rodrigues, and M.~J.~S.~Houndjo, Phys.\ Lett.\ B {\bf 715}, 241 (2012);
H.~R.~Kausar and I.~Noureen, Eur.\ Phys.\ J.\ C {\bf 74}, 2760 (2014).
\bibitem{Visser}
M. Visser, \emph{Lorentzian wormholes} (Springer-Verlag, New York, 1995).
\bibitem{LQG}
R.~Gambini and J.~Pullin, Phys.\ Rev.\ Lett.\  {\bf 110}, 211301 (2013).
\bibitem{SD}
H.~Gomes, Class.\ Quant.\ Grav.\  {\bf 31},  085008 (2014);
H.~Gomes and G.~Herczeg, Class.\ Quant.\ Grav.\  {\bf 31},  175014 (2014),
\bibitem{BroFab}
K. A. Bronnikov and J. C. Fabris, Phys. Rev. Lett. \textbf{96}, 251101 (2006).
\bibitem{fluids1}
R.~Shaikh, Phys.\ Rev.\ D {\bf 92},  024015 (2015).
\bibitem{fluids2}
T.~Harko, F.~S.~N.~Lobo, M.~K.~Mak, and S.~V.~Sushkov, Mod.\ Phys.\ Lett.\ A {\bf 30}, 1550190 (2015).
\bibitem{Tamang}
A.~Tamang, A.~A.~Potapov, R.~Lukmanova, R.~Izmailov, and K.~K.~Nandi, Class.\ Quant.\ Grav.\  {\bf 32},  235028 (2015).
\bibitem{Universe}
G. J. Olmo and D. Rubiera-Garcia, Universe  {\bf 1}, 173 (2015).
\bibitem{OlmoBook}
G.~J.~Olmo,  Springer Proc.\ Phys.\  {\bf 176},  183 (2016).
\bibitem{Chandra} S. Chandrasekhar, \emph{The mathematical theory of black holes} (Oxford University Press, New
York, 1992).
\bibitem{bcor16}
G. J. Olmo and D. Rubiera-Garcia,  Phys. Rev. D \textbf{84},  124059 (2011);
C. Bambi, A. Cardenas-Avendano, G. J. Olmo, and D. Rubiera-Garcia, Phys. Rev. D \textbf{93}, 064016 (2016).
\bibitem{dr}
J. Diaz-Alonso and D. Rubiera-Garcia, Phys. Rev. D \textbf{81}, 064021 (2010);
\textbf{82}, 085024 (2010);
Gen. Rel. Grav. \textbf{45}, 1901 (2013).
\bibitem{AdScore}
J. Bardeen, in Proceedings of GR5. Tiflis, U.S.S.R. (1968);
I.~Dymnikova,  Class.\ Quant.\ Grav.\  {\bf 21}, 4417 (2004);
S.~A.~Hayward, Phys.\ Rev.\ Lett.\  {\bf 96},  031103 (2006);
J.~P.~S.~Lemos and V.~T.~Zanchin, Phys.\ Rev.\ D {\bf 83},  124005 (2011);
I.~Dymnikova and E.~Galaktionov,  Class.\ Quant.\ Grav.\  {\bf 33},   145010 (2016);
E.~Spallucci and A.~Smailagic, arXiv:1701.04592 [hep-th].
\bibitem{sw}
G. J. Olmo, D. Rubiera-Garcia, and A. Sanchez-Puente, Eur. Phys. J. C \textbf{76}, 143 (2016).
\bibitem{congruences}
G. J. Olmo, D. Rubiera-Garcia, and A. Sanchez-Puente, Class. Quant. Grav. \textbf{33},  115007 (2016).
\bibitem{BIreg}
G.~J.~Olmo, D.~Rubiera-Garcia and A.~Sanchez-Puente, Phys.\ Rev.\ D {\bf 92} 044047 (2015)
\bibitem{hd}
D. Bazeia, L. Losano, G. J. Olmo, D. Rubiera-Garcia, and A. Sanchez-Puente, Phys. Rev. D \textbf{92},  044018 (2015).
\bibitem{3D}
D. Bazeia, L. Losano, G. J. Olmo, and D. Rubiera-Garcia, Class. Quant. Grav. \textbf{34}, 045006 (2017).
\bibitem{grapheneWH}
J. Gonzalez and J. Herrero, Nucl. Phys. B \textbf{825}, 426 (2010).
\bibitem{crystals}
F.~S.~N.~Lobo, G.~J.~Olmo and D.~Rubiera-Garcia, Phys.\ Rev.\ D {\bf 91}, 124001 (2015);
G.~J.~Olmo and D.~Rubiera-Garcia, Int.\ J.\ Mod.\ Phys.\ D {\bf 24}, 1542013 (2015).

\end{thebibliography}
\end{document}